\documentclass[10pt,twocolumn,twoside]{IEEEtran}
\usepackage{lineno,hyperref}
\usepackage{setspace}
\usepackage{amsmath}
\usepackage{graphicx}
\usepackage{algorithm}
\usepackage{geometry}
\usepackage{color}
\usepackage{amsmath}
\usepackage{amssymb}
\usepackage{amsfonts}
\usepackage{caption}
\usepackage{subcaption}
\usepackage{algpseudocode}
\usepackage{amsmath}
\usepackage{cases}
\usepackage{multirow}
\usepackage{etoolbox}
\usepackage{adjustbox}
\hypersetup{colorlinks=true,linkcolor=black}

\usepackage{stfloats}
  \modulolinenumbers[5]

\newtheorem{lemma}{Lemma}
\BeforeBeginEnvironment{lemma}{\noindent}
\newtheorem{theorem}{Theorem}

\graphicspath{{./plots/}, {./figures/}}

 \IEEEoverridecommandlockouts
\makeatletter
  \def\vhrulefill#1{\leavevmode\leaders\hrule\@height#1\hfill \kern\z@}
\makeatother

\graphicspath{{./plots/}, {./figures/}}
 
\begin{document}
\vspace{-0.1in}\title{ Spectral-Energy Efficiency Trade-off-based Beamforming Design for MISO   Non-Orthogonal Multiple  Access   Systems}
\author{Haitham Al-Obiedollah,   \IEEEmembership{ Student Member, IEEE}, Kanapathippillai Cumanan, \IEEEmembership{Senior Member, IEEE},  Jeyarajan Thiyagalingam,  \IEEEmembership{ Senior Member, IEEE},  Jie Tang, \IEEEmembership{ Senior Member, IEEE},  Alister G. Burr,     \IEEEmembership{ Senior Member, IEEE}, Zhiguo Ding,  \IEEEmembership{ Fellow, IEEE},  and Octavia A. Dobre, \IEEEmembership{ Fellow, IEEE} 
 \vspace{-0.4in}    
\thanks{H. Al-Obiedollah is with the Electrical Engineering Department, The Hashemite University, Zarqa, Jordan. (Email: haithamm@hu.edu.jo.)}

\thanks {K. Cumanan  and A. G. Burr are with the Department of Electronic Engineering, University of York, York, YO10 5DD, UK. (Email:   \{kanapathippillai.cumanan, alister.burr\}@york.ac.uk.)}
\thanks{J. Thiyagalingam is with the Scientific Computing Department
   of Rutherford Appleton Laboratory, Science and Technology
   Facilities Council, Harwell Campus, Oxford, UK. (Email:
   t.jeyan@stfc.ac.uk).}
   
\thanks{J. Tang is with the School of Electronic and Information
Engineering, South China University of Technology, Guangzhou, China. (email:
eejtang@scut.edu.cn).}
\thanks{Z. Ding is with the  School of Electrical and Electronic Engineering, The University of Manchester,  
 Manchester, UK.(Email: zhiguo.ding@manchester.ac.uk).}
 \thanks {O. A. Dobre is with the Department of Electrical and Computer
Engineering, Memorial University, St. John’s, NL A1B 3X5, Canada (email:
odobre@mun.ca).}
 
\thanks { The work of H. Al-Obiedollah was supported by the Hashemite University, Zarqa, Jordan. The work of Z. Ding was supported by the UK EPSRC under grant number EP/P009719/2.  The work of O. A. Dobre was supported by the Natural Sciences and Engineering Research Council of Canada (NSERC), through its Discovery program.}
}
 \maketitle

\begin{abstract}
Energy efficiency (EE) and spectral efficiency (SE) are two of the key performance metrics in future wireless networks, covering both design and operational requirements. {  For previous conventional resource allocation techniques,}   these    two  performance metrics have been considered in isolation, 
resulting in severe performance degradation in either of  these  metrics. Motivated by this   problem,    in this paper, we propose  a   novel  beamforming design that jointly considers the   trade-off between    the two  performance metrics  in a multiple-input single-output non-orthogonal multiple access system. In particular, we
formulate  a  joint SE-EE based design as a multi-objective optimization (MOO) problem to achieve
a good trade-off between    the   two  performance metrics. However, this MOO problem is not mathematically tractable and,   thus,      it is  difficult to determine a feasible solution due to the conflicting
objectives, where both need to be simultaneously optimized. To overcome this issue, we exploit a priori articulation scheme combined with the weighted sum approach. Using this, we reformulate the original MOO problem as a conventional single objective optimization (SOO) problem.  In doing so,  we develop an iterative algorithm to solve this non-convex SOO problem using the sequential convex approximation technique. Simulation results are provided to demonstrate the advantages and effectiveness of the proposed approach over the available beamforming designs.
  
\end{abstract}

\IEEEpeerreviewmaketitle
\section{Introduction}
Over recent years, extensive research efforts have been devoted to the  practical implementations of new disruptive technologies for   the fifth generation (5G)  and beyond wireless networks~\cite{book}. The unexpected exponential growth in the number of connected devices and the unprecedented  requirements of higher data rates,  low latency and  ultra reliability are of major concerns in future wireless networks~\cite{book}.   However, these demanding requirements are  difficult to meet or almost impossible to achieve  without  an enormous  power consumption, which is not only unacceptable due to  undesirable impacts on the natural environmental~\cite{green}, but also financially unaffordable.   Therefore,  it is important to  consider both as performance metrics,  the spectral efficiency (SE) and energy efficiency (EE), simultaneously.  SE is defined as the ratio between the achieved rate and the available bandwidth, whereas  EE is defined  as the ratio between the achieved sum rate in the system and the total required  power to achieve this  sum rate~\cite{ee1}.

\indent Different disruptive  technologies, including  massive multiple-input multiple-output (MIMO)~\cite{massive1} \cite{bashar2019energy}, millimeter-wave (mmWave) \cite{book}   \cite{zhu2019joint} \cite{xiao2018joint}, and   non-orthogonal multiple access (NOMA) techniques \cite{noma2} have been proposed  to meet the stringent design and operational requirements surrounding EE and SE.  In particular,   NOMA  has been envisioned as one of the key techniques for significantly improving the SE while providing massive connectivity to support the Internet-of-Things (IoT) in   5G   and beyond wireless networks \cite{fnoma}. Unlike the conventional orthogonal multiple access  (OMA) schemes, the users in NOMA  can be   served within  the same   resource blocks such as time, frequency, and code without any orthogonal divisions between them, by exploiting the power-domain multiplexing~\cite{noma5}. For instance, superposition coding (SC) is utilized at the base station to encode the transmit signals of multiple users with different transmit power levels~\cite{noma3} \cite{cuma9}. At the receiver end,  successive interference cancellation (SIC) is employed at the  strong users (i.e., the users with stronger/better channel gains)   to detect and remove the interference caused by  signals intended to  the weaker users    prior to decoding their own desired signals~\cite{noma2}. 

  To exploit  different potential benefits, NOMA has been recently integrated   with different technologies such as cognitive radio (CR) \cite{crnoma}, millimeter-wave    \cite{ref1}   \cite{ref3},   multiple-antenna  techniques  \cite{mimonom2}, \cite{misonoma}, \cite{zhang2019} and conventional OMA techniques \cite{xin}.  In particular, the combination of NOMA
with spatial domain multiple access (SDMA) offered by multiple-antenna  can provide additional benefits, by  jointly utilizing both the spatial and power domains.   The notion of joint utilization of multiple domains helps meeting the 
 demanding requirements in future wireless networks, particularly when 
  compared against   conventional stand alone SDMA techniques  \cite{mimonom2}, \cite{mimonoma}.  For instance,  incorporating  NOMA with the multiple-antenna techniques was considered in~\cite{mimonoma}. Another example is multiple-input single-output  (MISO)-NOMA~ \cite{sun2018robust} \cite{zhou2018artificial} \cite{zhou2017robust},    which  can be classified into two main categories:   beamformer-based      and   cluster-based MISO-NOMA schemes \cite{liu2018non} \cite{alavi2017outage}. 
 In this paper, we focus on    the   beamformer-based MISO-NOMA scheme,  where each user is served by a single beamforming vector.  For the sake of  notational  simplicity,    the beamformer-based MISO-NOMA  is  referred  to as MISO-NOMA throughout this paper.

 
  One of  the  conventional  beamforming designs   developed for MISO-NOMA systems in the literature considers SE  as a   performance metric with  the sum rate maximization (SRM) problem~\cite{hanif}. This SRM design is developed not only at the cost of the exponential increase in the available power, but also with significant loss in the EE performance.     In fact, with the     unprecedented  growth in the number of 
mobile devices and volume of mobile data traffic in future wireless networks,    EE  becomes   a prominent performance metric.   This is primarily
due to the fact that EE  has the potential   to   achieve  a good balance between    the  transmit power consumption and     system throughput \cite{ee1}. To overcome the EE degradation associated with the SRM design, we have proposed
a global EE maximization (GEE-Max) 
 design in our previous work       \cite{haitham} to maximize the overall EE of the system.      However, a major drawback of such a design   is that the base station does not have the flexibility to utilize the available power resources after achieving the maximum EE with its green power. In fact, this limitation becomes   an important issue that needs to be addressed   in    some of the scenarios, such as base station being powered by renewable energy sources
  \cite{haitham,wcnc1}.  Therefore,    the trade-off  between SE and EE  motivates one to explore novel design approaches for beamforming so that a good balance between SE and EE performance metrics can be achieved.  Furthermore, this  joint SE-EE design provides flexibility to  the base station to adapt the beamforming design by taking  the instantaneous transmission  conditions and the different requirements of the system   into consideration.    In particular, the  practical applications of the proposed SE-EE trade-off design  can be  summarized as follows:

  \begin{itemize}
  \item    Base stations with hybrid power   sources  are expected to play a crucial role on the deployments  of 5G and beyond wireless networks \cite{energyhybrid}. These hybrid base stations are powered by either non-renewable energy  sources   such as  diesel generators,   or renewable energy  sources  such as   photovoltaic panels and wind turbines to provide the communication services \cite{wang2012survey}, alternately. For such hybrid base stations, the priority to choose either    EE or SE depends on the available energy  source, i.e., if the base station utilizes a renewable energy resource, then the importance of  EE becomes less than that of   SE, and vice-versa for non-renewable energy  sources.  Hence,     an  SE-EE trade-off based design offers   flexibility  to the base station to switch between different design criteria based on the available energy  source. 
  \item  Furthermore, some resource allocation techniques aim  to maximize EE with an  SE constraint  \cite{xiong2011energy} \cite{alavi2019}. However, this design  limits the performance of either SE or EE due to its inflexibility \cite{eese}.  Hence, the SE-EE trade-off based design has a potential capability to  achieve  a good balance between these conflicting performance metrics, especially in some practical applications where both SE and EE  have similar importance.
    \end{itemize}

The joint SE and EE-based design can be developed by   formulating  a multi-objective optimization (MOO) problem with      these    two  performance metrics    in the  multi-objective function. In contrast to a unique global optimal solution in the conventional single objective optimization (SOO) problems, the MOO problems have many Pareto-optimal solutions which would yield a better performance in one of the multiple objectives  \cite{MO1}, \cite{MO2}. However, the required Pareto-optimal solution will be determined  based on the relative importance of each objectives in the overall problem. Therefore,  the   decision maker (DM) (i.e., the base station in our scenario) has to firstly  articulate the weights of  each  objective  prior to evaluate the solution for the MOO problem, which is referred as  a priori articulation in the literature~\cite{MO3}.  Then, those multiple objectives are converted   into    a single objective  function known as the utility  function to represent the corresponding  multi-objective functions \cite{wcnc2}.  In particular,    many utility functions  have been considered in the literature for different MOO problems,  including the weighted-sum \cite{MO2}, the weighted-product, and the weighted max-min function \cite{MO1}. As the MOO problems have different Pareto-optimal solutions,  the DM chooses the best trade-off solutions (Pareto-optimal solutions) \cite{MO1}.  It is worth mentioning that the SE-EE trade-off designs have been   considered in the wireless communications literature. For example,  an SE-EE design for a point-to-point communication link is considered in~\cite{eese}. Furthermore,  a generalized framework   for the SE-EE trade-off  was investigated for orthogonal frequency-division multiplexing (OFDM) in  \cite{eese1}.    Additionally, a multi-objective optimization approach is considered for link adaptation in an OFDM-based cognitive system in \cite{octavia}, where throughput and transmit power are simultaneously optimized.   A number of    MOO-based resource allocation techniques can be found in   \cite{xiong2011energy},  \cite{zhang2015energy}-\cite{hong2013energy}.

\subsection{Contributions}

  Motivated by the   importance of both   key performance metrics SE and EE in 5G and beyond wireless  networks \cite{book},  \cite{itu12}, and to overcome 
the limitations associated with  the conventional  GEE-Max  and   SRM  designs   \cite{haitham}-\cite{hanif},    in this paper we propose an SE-EE trade-off based design for an  MISO-NOMA system.   Unlike the conventional designs, this  SE-EE design optimizes SE and EE simultaneously to achieve a good balance between these conflicting performance metrics.  In doing that, we make the following key contributions:

\begin{itemize}
 
\item We formulate the overall SE-EE design as a MOO problem. Although this renders the overall problem as a challenging form, where direct approaches for obtaining a feasible solution are inherently difficult, it offers an avenue for achieving a good balance between SE and EE. This approach is radically different    from  the    conventional approaches, such as those outlined in~\cite{misonoma,hanif,haitham}, and~\cite{fayzeh}. More specifically, conventional optimization techniques, which are often employed in the context of  the SOO problems, cannot  directly  be applied to solve this  MOO problem.  Our design approach provides a  generic framework, where  the GEE max and SE max designs can be considered as   special cases by setting appropriate weight factors;

\item We provide  an algorithm  to solve this non-trivial MOO problem. This algorithm   utilizes  a priori articulation method combined with weighted-sum utility function  to    recast the MOO problem into a form of an SOO problem \cite{MO1}-\cite{MO3};  

\item We prove that solving the SOO problem provides a Pareto-optimal solution to the original MOO problem. In particular, the  sequential convex approximation (SCA) is exploited in the context of  handling the non-convexity of the SOO  problem.
 

\end{itemize}
 

The rest of the paper  is organized as follows. In Section~\ref{sec2}, the system model and the  problem formulation   are introduced to represent the SE-EE trade-off-based design. Section~\ref{sec3} presents    the proposed techniques to tackle   the SE-EE trade-off-based design. To verify the proposed beamforming design, numerical results are provided in Section~\ref{sec4}, where the performance of the  proposed beamforming design  is  compared with  that of the conventional beamforming  design criteria. Finally, conclusions are drawn in  Section~\ref{sec5}.

\subsection{Notations}
We use lower case boldface letters for vectors and upper case  boldface letters for matrices.  $(\cdot )^H $   denotes complex conjugate transpose,   and  $\Re(\cdot)$ and $\Im(\cdot)$  stand for  real and imaginary parts of a complex number, respectively. The symbols $\mathbb{C}^{N}$ and $\mathbb{R}^{N}$ denote  $N$-dimensional  complex and real spaces, respectively.        $||\cdot||_2 $  and $| \cdot| $ represent  the Euclidean norm of  a vector and the absolute value of a complex number, respectively.   $\mathbf{x} \succ0$ means that all the elements in the  vector  $\mathbf{x}$ are greater than zero.

\section{System Model And Problem Formulation}\label{sec2}
\subsection{System Model}
We consider a downlink transmission of a  MISO-NOMA system with $K$ single-antenna users in which    a    base station  equipped with  $N$   antennas simultaneously transmits to these $K$ users.  The transmit signal from the base station is  given by
 \begin{equation}\label{1}
\mathbf{x}=\sum_{j=1 }^{K} \mathbf{w}_j s_j,
\end{equation}
where  $s_j$ and $\mathbf{w}_j$   $ \in$ $\mathbb{C}^{N\times1}$  represent  the symbol intended to  the $j^{th}$ user, and the corresponding beamforming vector, respectively. It is assumed  that these symbols (i.e.,  $s_j, \forall j$) are independent and with unity power.   In addition,    digital beamforming is considered;  hence,   each user is served with a dedicated beamforming vector. As a result, we do not impose any constraint on the relationship between $K$ and $N$ and the proposed design is valid for any number of antennas and users.  
The received  signal at the $ i^{th}$ user can be written as
\begin{equation}\label{2}
y_i=\sum_{j=1 }^{K}\mathbf{h}_i^H \mathbf{w}_j s_j +n_i, 
\end{equation}
where   $n_i $ represents the zero-mean additive white Gaussian noise (AWGN) with variance $\sigma_i^2$, while $\mathbf{h}_i$  $ \in$ $\mathbb{C}^{N\times1}$ denotes the vector that contains the channel coefficients between the base station and the   $ i^{th}$ user.      Furthermore, we assume   frequency-flat channel conditions, and     the  channel coefficients can be modelled as 
$$ \mathbf{h}_i= \sqrt{d_i^{-\kappa}}\mathbf{g}_i,$$ where $ \kappa$ and $\mathbf{g}_i$  are the path loss exponent, and the small scale fading, respectively, whereas $d_i$ represents the distance between the $i^{th}$ user and the base station in meters.    We    consider  that    perfect channel state information (CSI) of the   users is available at the base station. \\

In the downlink power-domain  NOMA, the  power levels   are assigned to the users   based on their channel strengths such that  the allocated power levels   are  inversely  proportional to the   channel strengths of the users \cite{fnoma} \cite{noma2}.  Furthermore, the stronger users (i.e., users with higher channel strengths) perform   SIC  by firstly decoding  the signals intended to the users with weaker channel conditions, and then subtracting the decoded signals  prior of decoding their own  signals \cite{noma2} \cite{cuma2}. The weaker users detect  their   signals by  treating the interference caused by the signals intended to the stronger users as  noise \cite{noma4}. Hence, user ordering plays a crucial role in  power allocations, users' SIC capability, and the overall performance of the the NOMA systems. However, the optimal ordering could be determined  through performing an exhaustive search among all the user ordering possibilities, which is not practical to implement, especially in  dense networks. Therefore,  we consider the first user   (U$_1$) as the strongest user in the cell, whereas the   U$_K$  is the weakest user based on the following channel conditions:
\begin{equation}
 \underbrace{\vert\vert \mathbf{h}_K \vert\vert^2}_{\text{Weakest }} \leq \vert\vert \mathbf{h}_{K-1} \vert\vert^2\leq \cdots \leq  \underbrace{\vert\vert \mathbf{h}_1\vert \vert^2}_{\text{Strongest}}. 
\end{equation}\\

 Based on this user ordering,    to  ensure that the power allocated to each user in the system is inversely proportional to its channel gain, and to successfully implement SIC  at the stronger users \cite{hanif}, the following conditions should be satisfied with the beamforming design \cite{haitham}: 
\begin{equation}\label{channels}
 \vert \mathbf{h}_i^H\mathbf{w}_K \vert^2 \geq \cdots \geq\vert \mathbf{h}_i^H \mathbf{w}_{1}\vert^2, \forall  i \in \mathcal{K}.
\end{equation} 

 It is worthy to point out that   some work  in the context of downlink NOMA transmission   literature    assumes that NOMA transmission  can be achieved without including the constraint in  (\ref{channels}), such as in \cite{mojtaba}. However,    this power allocation constraint facilitates the design SIC orders, and   has been assumed in most of the works in the literature to ensure successful implementation of SIC.

Therefore, the  received signal at U$_i$  after performing  SIC  is written as 
\begin{equation}\label{3}
\overset{*}{y_i}=\underbrace{\mathbf{h}_i^H\mathbf{w}_is_i}_{\text{Intended signal }}+\underbrace{{\sum_{j=1 }^{i-1} \mathbf{h}_i^H \mathbf{w}_j s_j}}_{\text{Interference }} +\underbrace{n_i}_{\text{Noise }},   \forall  i \in \mathcal{K}, 
\end{equation}
where $\mathcal{K} \overset{\bigtriangleup}{=}\{1, \cdots,K\}$. Note that the interference caused by  U$_{i+1},\cdots, $U$_{K}$ is removed through  SIC.
Furthermore,  U$_k$     has the capability to  decode  the message of U$_i$ ($k\leq i$) with {   signal-to-interference and noise ratio (SINR) that  can be written as  
\begin{equation} 
\text{SINR}^{(i)}_k= \frac{\vert \mathbf{h}_k^H\mathbf{w}_i \vert^2}{\sum_{j=1 }^{i-1}\vert \mathbf{h}_k^H \mathbf{w}_j \vert^2+\sigma_k^2}, \forall  i \in \mathcal{K}, k\leq i.
\end{equation}
  Now, with the assumption that $s_i$ is only decodable provided its SINR is higher than a  threshold denoted as SINR$^{th}$,    this explicitly requires that decoding of $s_i$ at other stronger users should be also higher than this  threshold  \cite{hanif},  i.e., SINR$_k^{(i)} \geq $    
  SINR$^{th}$, $\forall k=1,2,\cdots, i$.   Based on  this argument, the definition of SINR$_i$   should   take  into account the  decoding of  $s_i$ at the stronger users in order to align with the basic principle of NOMA, namely   SIC. Based on this requirement, the achievable SINR can be defined as follows: 
\begin{equation}\label{sinr}
\text{SINR}_i= \text{min} ({\text{SINR}^{(i)}_{1},\text{SINR}^{(i)}_{2},\cdots,\text{SINR}^{(i)}_{i}}), \forall  i \in \mathcal{K}.
\end{equation}
Based on the above discussion,  the  achieved  rate at   U$_i$  can be defined as \cite{hanif}
\begin{equation}\label{rate}
R_i= \text{min} ({R^{(i)}_1,R^{(i)}_{2},R^{(i)}_{3}, \cdots,R^{(i)}_{i}}), \forall  i \in \mathcal{K}. 
\end{equation}
Note that  $R^{(i)}_k$ is the rate of decoding $s_i$ at U$_k$, and it is given as }
 
\begin{equation} 
R^{(i)}_k=B_w\log_2\Bigg(1+\frac{\vert \mathbf{h}_k^H\mathbf{w}_i \vert^2}{\sum_{j=1 }^{i-1}\vert \mathbf{h}_k^H \mathbf{w}_j \vert^2+\sigma_k^2}\Bigg), \forall  i \in \mathcal{K},   
\end{equation} 
where $B_w$ is the available bandwidth,  set to be one in  this analysis.

\indent The global energy efficiency (GEE)\footnote{GEE and EE   carry  the same meaning throughout this paper.} of the system is defined as the ratio between  the achieved sum rate of the system and  the total power required  to achieve  this rate  (bits/Joules) \cite{ee1}, and is expressed as 
\begin{equation}\label{gee}
\text{EE}=\text{GEE}=\frac{\sum_{i=1}^{K}R_i}{\frac{1}{\epsilon_0}P_t +P_{l}},
\end{equation}
where $\epsilon_0$ denotes the power amplifiers efficiency at the base station. Furthermore,   $P_l$ and   $P_t$ represent the power losses at the base station and the  transmit power, respectively.  Note that   $P_t$   should be less than the the available power budget at the base station ($P_{ava}$) which  can be expressed as the following constraint: 
\begin{equation}\label{gee}
P_t=\sum_{i=1}^{K}||\mathbf{w}_i||_2^2\leq P_{ava},
\end{equation}  
 For the MISO-NOMA system considered in this paper,
the GEE maximization  (GEE-Max)   design can be  formulated into the following optimization problem \cite{haitham}:

\begin{subequations}\label{ee_max1}
\begin{align}
 {OP_{EE }}:
& \underset{\{\mathbf{w}_i\}_{i=1}^{K}}{\text{max}}
& &\text{GEE} \\
& \text{subject to}
& & \sum_{i=1}^{K}\vert \vert \mathbf{w}_i \vert \vert ^2_2\leq P_{ava}, \label{jo1}\\
& &&  R_i \geq R^{th}_i, \forall  i \in \mathcal{K},\label{rate_co1}\\
&&&   (\ref{channels}).\label{cha11}
\end{align}
\end{subequations}
The constraint in  (\ref{rate_co1}) ensures that each user can achieve minimum predefined threshold rate ($ R^{th}_i$) which is referred to as minimum rate constraint. Furthermore,  the constraint in (\ref{cha11}) facilitates the successful implementation of   SIC which is referred to as the SIC constraint throughout this paper. It is worth mentioning that the GEE-Max problem ${OP_{EE }}$ is solved in \cite{haitham} using the SCA technique and the Dinkelbach's algorithm. In particular, the maximum GEE is achieved with a  certain available power which is known as  the green power in the literature \cite{zhang2017secure} \cite{oct4}. Beyond this green power, both GEE and the achieved sum  rate saturate \cite{islam2018}.\\
 \indent   Now, we formulate the SE maximization (SE-Max) problem for the MISO-NOMA system defined in this paper, as follows \cite{hanif}:   
 \begin{subequations}
  \label{SE_max}
\begin{align}
 {OP_{SE }}:
 &\underset{\{\mathbf{w}_i\}_{i=1}^{K}}{\text{max }}
& &\text{SE} \\
& \text{subject to}
& & \sum_{i=1}^{K}\vert \vert \mathbf{w}_i \vert \vert ^2_2\leq P_{ava} \label{jo1}\\
&&&  (\ref{channels})\label{cha1}, 
\end{align}
\end{subequations}
  where $\text{SE}=\frac{\sum_{j=1}^{K}R_j}{B_w}.$  Note that   the  SE-Max  problem is equivalent to the conventional SRM with the assumption of unit bandwidth, i.e., $B_w=1$. Hence,  without loss of generality, SE-Max and SRM refer to the same problem throughout this paper. In particular,  this SE-Max   problem is   solved for the MISO-NOMA system in \cite{hanif}. Note that  if  the minimum rate constraint in (\ref{rate_co1}) is added to  the original SE-Max $OP_{SE}$, then, the modified   SE-Max problem will be referred as  SE-Max (min-rate)  in this paper. It is obvious that  these conventional  SE-Max and GEE-Max   designs  maximize  either  EE or  SE  individually, without jointly considering them   to achieve a trade-off  between these performance metrics. In the following subsection,  we  develop a joint SE-EE trade-off   design.   
\subsection{Problem Formulation}
For   notation simplicity, we represent SE and  EE (i.e., GEE)  by the functions $f_1\left(\{\mathbf{w}_i\}_{i=1}^{K}\right)$ and $f_2\left(\{\mathbf{w}_i\}_{i=1}^{K}\right)$, respectively. In particular, we aim to develop a beamforming design that can \textit{jointly} maximize these  performance metrics (i.e., max $f_1\left(\{\mathbf{w}_i\}_{i=1}^{K}\right)~ \text{and}~ f_2\left(\{\mathbf{w}_i\}_{i=1}^{K}\right)$ with the given set of  constraints.  Therefore, the beamforming vectors that can achieve a trade-off between the conflicting SE and EE metrics in the considered  MISO-NOMA system can be formulated into the following MOO problem:
\begin{subequations}\label{5}
\begin{align}
{OP }:
& \underset{\{\mathbf{w}_i\}_{i=1}^{K}}{\text{max}}  
& & \mathbf{f}\left(\{\mathbf{w}_i\}_{i=1}^{K}\right)  \\
 & \text{subject to} 
& & \sum_{i=1}^{K}\vert \vert \mathbf{w}_i \vert \vert ^2_2\leq P_{ava},\label{oper} \\
& &&  R_i \geq R^{th}_i, \forall i \in \mathcal{K},\label{rate_cco}\\
&&&  (\ref{channels}).\label{ccha}
\end{align}
\end{subequations}
 Note that   the objective vector  $\mathbf{f}\left(\{\mathbf{w}_i\}_{i=1}^{K}\right) $ consists of the   SE and GEE  functions, such that $\mathbf{f}\left(\{\mathbf{w}_i\}_{i=1}^{K}\right)=   \left[  f_1\left(\{\mathbf{w}_i\}_{i=1}^{K}\right),  f_2\left(\{\mathbf{w}_i\}_{i=1}^{K}\right) \right]$. It is obvious that there is no   global optimal solution that maximizes these conflicting  objectives in $OP$ \cite{MO1}. However, the MOO problem $OP$ searches for all possible  best trade-off solutions, which are  known as the  Pareto-optimal solutions in the literature \cite{MO3}.\\

\textbf{Definition 1. \cite{MO2} \cite{MO3}} A feasible solution ${\{\mathbf{w}_i^*\}_{i=1}^{K}}$ is defined as a Pareto-optimal solution if  there exists no other feasible solution  ${\{\mathbf{w}_i^{'}  \}_{i=1}^{K}}$ such that $\mathbf{f}\left(\{\mathbf{w}_i^{'}\}_{i=1}^{K}\right) \succ \mathbf{f}\left(\{\mathbf{w}_i^*\}_{i=1}^{K}\right)$. The set of all Pareto-optimal solutions are collectively defined as the Pareto front in the { literature} \cite{MO2}.\\

Therefore, our aim   is to find the set  of the feasible solutions that satisfy the Pareto-optimality conditions for this MOO problem.   However, due to the fact  that the original $OP$  problem might be infeasible with certain $P_{ava}$,  we carry out a feasibility check prior to solving it. The feasibility check and the proposed methodology to solve the problem   $OP$ are provided in the next section.

\section{Proposed Methodology}\label{sec3}
First, we carry out a feasibility check prior to solving the optimization problem $OP$. For infeasible problems,   we propose another beamforming design. The feasible   $OP$ is solved by reformulating it as SOO problem using a priori articulation technique combined with the weighted-sum approach.  Then,  the SCA technique is exploited  to tackle the non-convexity issue of the SOO problem. At the end of this section,  we provide some discussions on  the convergence  and the performance evaluation  of the proposed SCA algorithm to solve $OP$.
\subsection{Feasibility Check}
Firstly, it is worthy to mention  that the original optimization problem $OP$ turns out to be infeasible when the minimum rate requirements at each user cannot be met with the available   power budget at the base station (i.e., $P_{ava}$). Therefore, it is important to investigate the feasibility of the   original problem $OP$ prior to solving it. In particular, this feasibility check can be performed through evaluating the minimum   transmit power, referred as $P^*$, that is required to satisfy the minimum rate   and SIC constraints, in (\ref{rate_cco}) and (\ref{ccha}), respectively. This $P^*$ can be determined through solving the following   power minimization problem:

\begin{subequations}\label{min}
\begin{align}
{OP_P:~P^*}=
&\underset{\{\mathbf{w}_i\}_{i=1}^{K}}{\text{min}}
& &\!\!\!\!\!\!\!\!\!\!\!\! \sum_{i=1}^{K}\vert \vert \mathbf{w}_i \vert \vert ^2_2 \\
& \text{subject to}
& &  (\ref{rate_cco}),(\ref{ccha}). 
\end{align}
\end{subequations}

Note that the original  problem $OP$ can only be solved provided that $P^*\leq P_{ava}$, {and   is infeasible when $P^*$ is higher than the available power at the base station (i.e., $ P_{ava}$)}. To overcome this infeasibility,   an alternative   beamforming design can be considered to maximize  the sum rate with available power budget, as in $OP_{SE}$ defined in (\ref{SE_max}).  Without loss of generality, we assume that $OP$ is feasible (i.e., $P^*\leq P_{ava}$), and propose an effective approach to solve it in the following subsection.

\subsection{Proposed Methodology}
  As mentioned before, we first reformulate the original MOO problem $OP$ into a SOO form. Then, we employ the SCA technique to solve the   SOO problem. More details are provided in the following discussions.

\subsubsection{Single Objective Transformation}
First, we use  a priori articulation scheme where   the base station  determines the  relative importance of each objective function prior to determining the beamforming vectors based on the design requirements. In particular,  the weight factor $\alpha_i$ is assigned to the  $i^{th}$ objective function (i.e., $f_i (\{\mathbf{w}_i\}_{i=1}^{K}$)  to reflect its relative importance   on the overall design, such that $\sum_{i=1}^{2}\alpha_i=1$, $\alpha_i \in [0,1]$. Then,   the vector containing the objective functions in the original MOO problem (i.e., $OP$) {  is replaced with}  a single objective function known as the utility function in the literature.  Note that  the utility function is   a single-objective function that can alternatively
 represent   the original multi-objective function based on the importance of each objective function \cite{MO1}. There are several utility functions  available in the literature \cite{MO1} \cite{MO2} \cite{MO3};  we choose the weighted sum approach as it   provides the Pareto-optimal solution to the original problem,  as shown in Theorem 1. Based on the previous discussion, the SOO framework that represents the original MOO  problem  $OP$ can be formulated as follows:

 \begin{subequations}\label{6}
\begin{align}
{\overset{\sim}{OP} }:
&\underset{\{\mathbf{w}_i\}_{i=1}^{K}}{\text{max}}
& & f_{EE-SE}\left(\{\mathbf{w}_i\}_{i=1}^{K}\right)  \\
& \text{subject to}
& & \sum_{i=1}^{K}\vert \vert \mathbf{w}_i \vert \vert ^2_2\leq P_{ava}, \label{sara1} \\
& &&  R_i \geq  R^{th}_i, \forall i \in \mathcal{K},\label{rate_co}\\
&&&  (\ref{channels})\label{cha},
\end{align}
\end{subequations}
where $f_{EE-SE}\left(\{\mathbf{w}_i\}_{i=1}^{K}\right)=  \sum_{l=1}^{2} \alpha_l f_l^{Norm}\left(\{\mathbf{w}_i\}_{i=1}^{K}\right)$.  Note that  $f_1^{Norm}\left(\{\mathbf{w}_i\}_{i=1}^{K}\right) $ and  
 $f_2^{Norm}\left(\{\mathbf{w}_i\}_{i=1}^{K}\right)$ represent  the unit-less normalized
 version   of  $f_1 \left(\{\mathbf{w}_i\}_{i=1}^{K}\right)$ and $f_2\left(\{\mathbf{w}_i\}_{i=1}^{K}\right) $, respectively, which can be defined as
\begin{subequations}\label{7}
\begin{align} 
  f_1^{Norm}\left(\{\mathbf{w}_i\}_{i=1}^{K}\right) =\frac{ {f_1\left(\{\mathbf{w}_i\}_{i=1}^{K}\right)}}{{f_1}^*},\\ 
f_2^{Norm}\left(\{\mathbf{w}_i\}_{i=1}^{K}\right) =\frac{f_2\left(\{\mathbf{w}_i\}_{i=1}^{K}\right)}{f_2^*},     
\end{align}
\end{subequations}  
where  $f_1^*$ and  $f_2^*$ are the maximum values of SE and GEE, respectively. {  In particular, $f_1^*$ and  $f_2^*$   can be determined through solving $OP_{SE}$ and  $OP_{EE}$, respectively.}    
Note that the normalization of the  objectives { in (\ref{7})} is an important step   in the context of solving the original MOO problem  $OP$  due to   several  reasons. Firstly, it is obvious that the  performance metrics GEE and SE have different units.   Therefore,      an  addition of such (un-normalized) functions  is neither allowable nor defines
  any meaningful performance metric.    Secondly, as these two functions have completely different ranges, combining them   uisng  a weighted-sum utility function will certainly  degrade the achievable objective value of the function with lower range \cite{MO1}. Therefore, to treat both objective functions in a fair manner, we employ
a unitless normalization,   by dividing each  objective function  with its corresponding  
 optimal value.  With such a normalization, we obtain a non-dimensional objective function with an upper bound of one.  Note that    different    normalization (i.e., transformations) methods have been   considered   for MOO  problems in the literature \cite{eese}, \cite{MO1},   \cite{MO3}.  
For   notation simplicity, we use    $\alpha_2 =\alpha$  and   $\alpha_1 =1-\alpha$. To examine the Pareto-optimality of ${\overset{\sim}{OP} }$, we present the following theorem:
\begin{theorem}
The { solutions of the} weighted-sum SOO problem  in ${\overset{\sim}{OP} } $  provide  the  Pareto-optimal solutions for the original   MOO ${OP}$ problem. 
\end{theorem}
\textit{Proof}:  Please refer to    Appendix A.  \hfill$\blacksquare$\\

It is obvious that     ${\overset{\sim}{OP} }$ turns out to be  SE-Max (min-rate) when $\alpha=0$. Furthermore,  the problem becomes  GEE-Max  with $\alpha=1$. However, a good balance between the conflicting SE and EE  can be achieved through choosing an appropriate   $\alpha$ between 0 and 1.  To this end, we have transformed the original MOO problem $OP$ into a form of a SOO problem ${\overset{\sim}{OP} }$. However,   the   optimization problem  ${\overset{\sim}{OP} }$ cannot be directly solved due to the non-convexity nature of the objective function and the corresponding constraints.  To circumvent this non-convexity issue,  we propose an effective approach to solve $ {\overset{\sim}{OP} }$ in the next subsection.

 \subsubsection{Sequential Convex Approximation}
  The  SCA technique is an iterative approach to solve the original non-convex optimization problem by approximating the non-convex functions by lower-bounded convex functions \cite{sqa} \cite{harv}. In particular, the SCA technique has been employed to solve different non-convex  resource allocation problems in the literature \cite{hanif} \cite{haitham}. Similarly, we exploit the SCA technique to solve the  $ {\overset{\sim}{OP} }$  problem by approximating each non-convex term with a convex one. We start with the objective function by introducing two new slack variables  $\Gamma_1$ and $\Gamma_2$   such that 
\begin{subequations}\label{8}
\begin{align} 
  (1-\alpha)f_1^{Norm}\left(\{\mathbf{w}_i\}_{i=1}^{K}\right) \geq\Gamma_1,  \\ 
\alpha f_2^{Norm}\left(\{\mathbf{w}_i\}_{i=1}^{K}\right) \geq\Gamma_2.      
\end{align}
\end{subequations}

Based on these slack variables, the original   $ {\overset{\sim}{OP} }$  problem can be equivalently written as 
\begin{subequations}\label{9}
\begin{align}
&{\overset{\approx}{OP} }:\underset{ \Gamma_1, \Gamma_2,  \{\mathbf{w}_i\}_{i=1}^{K}}{\text{max}}
&&  \Gamma_1 +\Gamma_2 \\
& ~~~~~~~~~~~~~~\text{subject to} 
& & \sum_{i=1}^{K}\vert \vert \mathbf{w}_i \vert \vert ^2_2\leq P_{ava}, \label{power} \\
&  &&   R_i \geq R_{th},\forall i \in \mathcal{K},\label{bn1}\\
&&  &(\ref{channels}),\label{bn2}\\
&&  &\alpha f_2^{Norm}\left(\{\mathbf{w}_i\}_{i=1}^{K}\right) \geq\Gamma_2, \label{go}\\
&& &(1-\alpha) f_1^{Norm}\left(\{\mathbf{w}_i\}_{i=1}^{K}\right) \geq\Gamma_1. \label{moo4}
\end{align}
\end{subequations} 
It is obvious that the objective function in $ \overset{\approx}{OP}$   is a linear  function in terms of  $ \Gamma_1 $ and $\Gamma_2$. Furthermore,  the constraints are not convex and  we  handle these non-convexity issues as follows.  First, we look into the non-convexity of the constraint in  (\ref{moo4}) by 
 rewriting it as 
\begin{equation}\label{ccc}
\sum_{i=1}^{K}\log_2(1+\text{SINR}_i) \geq \frac{f_1^*}{(1-\alpha)}\Gamma_1.
\end{equation}
We handle  this non-convexity issue by introducing  new  slack variables $z_i, \rho_i $, such that
\begin{subequations}\label{slk}
\begin{align} 
& \log_2 (1+ \text{SINR}^{(i)}_{k})\geq \rho_i, \forall i \in \mathcal{K},k\leq i, \label{cc}\\
 &1+\text{SINR}^{(i)}_{k}\geq z_i,\forall i \in \mathcal{K},k\leq i.\label{kok}
\end{align}
\end{subequations} 
 Based on these multiple slack variables,  the constraint in (\ref{ccc}) can be equivalently written  as the following set of constraints:
 \begin{subnumcases}{(\ref{ccc})  \Leftrightarrow  }
    \sum_{i=1}^{K} \rho_i \geq  \frac{f_1^*}{(1-\alpha)}\Gamma_1, \label{fst}
   \\
  z_i \geq 2^{\rho_i}, \quad \forall i \in \mathcal{K}, \label{sec}
   \\
   (\ref{kok}).  \label{thd}
\end{subnumcases}
It is obvious that the inequalities (\ref{fst}) and (\ref{sec}) are convex constraints, whereas  the constraint in (\ref{thd}) remains still non-convex. Furthermore, we introduce another slack  variable  $ a_{i,k}$ { to convert it into a convex one as follows:}
\begin{subequations}\label{slk}
\begin{align} 
 {\vert \mathbf{h}_k^H\mathbf{w}_i \vert^2} \geq  {(z_i-1) a_{i,k}^2},\forall i \in \mathcal{K},  k\leq i,\label{10}\\ 
 a_{i,k}^2 \geq {\sum_{j=1 }^{i-1}\vert \mathbf{h}_k^H \mathbf{w}_j \vert^2+\sigma_k^2} ,\forall i \in \mathcal{K}, k\leq i.\label{11}
\end{align}
\end{subequations} 
  We handle the non-convexity issues in the constraint (\ref{10}) by   approximating $\vert \mathbf{h}_k^H\mathbf{w}_i \vert^2$ with  a lower  bound which is chosen to be $\Re({\mathbf{h}_k^H\mathbf{w}_i})$, such that  
\begin{equation}\label{neweq}
\vert \mathbf{h}_k^H\mathbf{w}_i \vert^2 \geq {\left(\Re({\mathbf{h}_k^H\mathbf{w}_i})\right)}^2, \forall k, \forall i.
\end{equation}    
\noindent Note that the constraint in (\ref{neweq})   is always held true  for any set of channel coefficients and beamforming vectors,  and thus,   is  not required to be included in the optimization problem. Now, we   take the square-root of both sides in (\ref{10}) after incorporating the new approximation in (\ref{neweq}). Next,   the right-hand side of this inequality can now be approximated with   linear   function using the  first-order Taylor series approximation. Based on that, the constraint in (\ref{10}) can be written in the following approximated convex form:
\begin{multline}\label{cuma4}
\Re({\mathbf{h}_k^H\mathbf{w}_i})\geq   \sqrt{( z_i^{(n)}-1)} a_{i,k}^{(n)}\\+0.5\frac{1}{\sqrt{( z_i^{(n)}-1)}} a_{i,k}^{(n)}( z_i- z_i^{(n)})\\+  \sqrt{( z_i^{(n)}-1)}(a_{i,k}-a_{i,k}^{(n)}), \forall i \in \mathcal{K}, k\leq i, 
\end{multline}
where $ a_{i,k}^{(n)}$ and $z_i^{(n)}$  represent the approximations of $ a_{i,k} $ and $z_i $ in the n$^{th}$ iteration, respectively.
However,  the constraint in  (\ref{11}) can be reformulated into    the following second-order cone (SOC) \cite{convex}: 
 \begin{equation}\label{cuma5}
 a_{i,k} \geq || \left[\mathbf{h}_k^H \mathbf{w}_{i-1} \cdots\mathbf{h}_k^H \mathbf{w}_1~\sigma_k\right]^T ||_2,\forall i \in \mathcal{K}, k\leq i. 
 \end{equation}
 \indent Next,   the non-convexity of the constraint in (\ref{go}) is tackled  by introducing a new slack variable $b$  such that 
 \begin{equation}
 \frac{\sum_{j=1}^{K}R_j}{\frac{1}{\in}P_t +P_{l}} \geq \frac{f_2^*}{\alpha}\frac{\Gamma_2 b^2}{b^2},
 \end{equation} 
hence, the constraint in (\ref{go}) can be split into the following two constraints:
\begin{subequations}\label{go1}
\begin{align} 
 \sum_{j=1}^{K}R_j  \geq  \frac{f_2^*}{\alpha} {\Gamma_2 b^2},\label{go2} \\
 {b^2} \geq {\frac{1}{\epsilon_0}\sum_{i=1}^{K}||\mathbf{w}_i||_2^2 +P_{l}}.\label{go3}
\end{align}
\end{subequations} 
To resolve the non-convexity issue  in  (\ref{go2}),  we   exploit the same approaches used to handle the constraint in  (\ref{ccc})    by introducing a set of   new slack variables,  $ r_i,\xi_{i,k}$, and   $\rho_i$,  such that
\begin{subequations}
\begin{align}
&R_i \geq \rho_i,\forall i \in \mathcal{K},\\
&\frac{\vert \mathbf{h}_k^H\mathbf{w}_i \vert^2}{\sum_{j=1 }^{i-1}\vert \mathbf{h}_k^H \mathbf{w}_j \vert^2+\sigma_k^2}\geq (r_i-1) \frac{\xi_{i,k}^2}{\xi_{i,k}^2},  \forall i \in \mathcal{K}, k\leq i.
\end{align}
\end{subequations}
 Based on these multiple slack variables, the constraint in  (\ref{go2}) can be approximated through   the following convex constraints:
\begin{multline}\label{slk}
\Re({\mathbf{h}_k^H\mathbf{w}_i})\geq   \sqrt{( r_i^{(n)}-1)} \xi_{i,k}^{(n)}\\+0.5\frac{1}{\sqrt{( r_i^{(n)}-1)}} \xi_{i,k}^{(n)}( r_i - r_i^{(n)})\\+   \sqrt{( r_i^{(n)}-1)}(\xi_{i,k} -\xi_{i,k}^{(n)}), \forall i \in \mathcal{K},k\leq i,
\end{multline}  
\begin{equation} \label{cuma1}
\xi_{i,k} \geq || \left[\mathbf{h}_k^H \mathbf{w}_1 \cdots \mathbf{h}_k^H \mathbf{w}_{i-1}~\sigma_k\right]^T ||_2,\forall i \in \mathcal{K},k\leq i, 
\end{equation}
\begin{equation}\label{s1c}
 r_i \geq 2^{ \rho_i}, \quad \forall i \in \mathcal{K},
\end{equation}  
 \begin{multline}\label{lara1} 
 \sum_{j=1}^{K}{ \rho_i}  \geq \\   \frac{f_2^*} {\alpha}\Big( \Gamma_2^{(n)} {b^2}^{(n)}+ 2b^{(n)}\Gamma_2^n(b -b^{(n)})  +  {b^2}^{(n)}( \Gamma_2-\Gamma_2^{(n)})\Big).
 \end{multline}
\\
Similar to the constraint in (\ref{11}),  the constraint in (\ref{go3}) can be cast as  the following SOC  constraint: 
\begin{equation}\label{haith}
b \geq \frac{1}{\sqrt{\epsilon_0}}|| \left[ ||\mathbf{w}_1||_2~||\mathbf{w}_2||_2 \cdots  ||\mathbf{w}_{K}||_2~\sqrt{P_l} \right]^T ||_2.  
 \end{equation}
 To this end, the non-convex constraint in (\ref{go}) is replaced with the following convex constraints: 
\begin{subnumcases}{\text{(\ref{go})}  \Leftrightarrow  }
    (\ref{slk}),  (\ref{cuma1}),  (\ref{s1c}),\\ (\ref{lara1}),(\ref{haith}).
\end{subnumcases} 
 \indent Next, the  non-convexity of the constraint in (\ref{bn2}) is handled    by replacing each term in the inequality by a linear term using the first-order Taylor series expansion, such that 
\begin{multline}\label{noora1}
\vert \mathbf{h}_k^H\mathbf{w}_i \vert^2  \geq |\vert \left[ \Re{\left(\mathbf{h}_k^H \mathbf{w}_i^{(n)}\right)} ~ \Im{(\mathbf{h}_k^H \mathbf{w}_i^{(n )})}\right]^T||^2 \\+2 \left[ \Re{(\mathbf{h}_k^H \mathbf{w}_i^{(n )})} ~ \Im{(\mathbf{h}_k^H \mathbf{w}_i^{(n )})}\right] \\   
 \big[(\Re{(\mathbf{h}_k^H \mathbf{w}_i )}-\Re{(\mathbf{h}_k^H \mathbf{w}_i^{(n)})})(\Im{(\mathbf{h}_k^H \mathbf{w}_i )}-\Im{(\mathbf{h}_k^H \mathbf{w}_i^{(n )})})\big]^T.  
\end{multline}
Note that the right-hand side of the inequality in (\ref{noora1}) is linear  in terms of  $\mathbf{w}_i$. Hence, each term in the constraint in (\ref{bn2})  is replaced by the right-side of  (\ref{noora1}). Finally, we consider the constraint in (\ref{bn1})  with the following  equivalent SINR constraint:
\begin{equation}\label{soc}
\frac{\vert \mathbf{h}_k^H\mathbf{w}_i \vert^2}{\sum_{j=1 }^{i-1}\vert \mathbf{h}_k^H \mathbf{w}_j \vert^2+\sigma_k^2}\geq  \eta^{th}_i,\forall i \in \mathcal{K},k\leq i,
\end{equation} 
where   $ \eta^{th}_i=2^{ R^{th}_i}-1 $. Furthermore,  the constraint in  (\ref{soc})  can be reformulated as the following SOC constraint:
\begin{multline}\label{const12}
\frac{1}{\sqrt{ \eta^{th}_i}}\Re({\mathbf{h}_k^H\mathbf{w}_i})\geq \vert \vert \left[ \mathbf{h}_k^H \mathbf{w}_{1}~\cdots \mathbf{h}_k^H \mathbf{w}_{i-1}~\sigma_k \right]^T \vert \vert_2 , \\ \forall i \in \mathcal{K},k\leq i.
\end{multline}
Based on these approximations,    the original non-convex optimization problem ${\overset{\sim}{OP} }$ can be reformulated as 
\begin{subequations}\label{20}
\begin{align}
& {\overset{\cong}{OP} }:\underset{\Psi }{\text{Maximize }}
&&  \Gamma_1  +\Gamma_2  \\
& \text{subject to} 
& & (\ref{bn2})\footnotemark[1],(\ref{sec}),(\ref{s1c}),\\
&  &&  (\ref{cuma4}),(\ref{cuma5}),(\ref{slk}),(\ref{cuma1}),(\ref{const12}),\label{21}\\
&&& (\ref{power}),  (\ref{fst}), (\ref{lara1}), (\ref{haith}),
\end{align}
\end{subequations} 
\noindent where $\Psi$ consists of all the variables involved in this design, which can be expressed  as \\
$  \Psi=\{\mathbf{w}_i,r_i,b,\Gamma_1,\Gamma_2,z_i, \xi_{i,k},a_{i,k},\rho_i \}_{i=1}^{K}. $ 
\footnotetext[1]{Replace (\ref{noora1}) instead of each term in the inequality.}

Note that    the relationship between SE and EE basically shows   two   different trends with the available power. In the first trend, both   SE and   EE increase with the available power and this trend continues until the available power reaches the green power. Once  the available power exceeds the green power, both   SE and   EE show the conflicting nature with the available power, which leads to   the  second trend. In order to    shed more light  on these trends, we provide the following lemma: 
\begin{lemma}   The SE-EE optimization problem $ {\overset{\sim}{OP} }$ provides the same   solution   with different weight factors $\{\alpha_l\}_{l=1}^{2}$ 	when the available power  $P_{ava}$ is less than the green power (i.e., $ P_{ava} \leq\text{ green power} )$. 
\end{lemma}
\textit{Proof}:  Please refer to    Appendix B. \hfill$\blacksquare$\\
It is worth mentioning that  the solution of  $ {\overset{\cong}{OP} }$  requires an appropriate selection of the initial parameters (i.e., $\Psi^{(0)}$). As this is an iterative approach, it is important to provide discussion on the initial conditions and the convergence of the proposed algorithm, which are presented in the following  subsection. 
{ \subsection{Initial Conditions, Convergence Analysis,   Performance Evaluation, and Complexity Analysis}}
\subsubsection{Initial Conditions}
Firstly, it is crucial to choose an appropriate set    $ \Psi^{(0)}$ to ensure the feasibility of the problem in the first iteration of the algorithm \cite{xin}. In particular, we   choose a set of feasible  beamforming vectors which can satisfy all the constraints in the approximated  problem $ {\overset{\cong}{OP} }$. Then, we determine all required slack variables in $ {\overset{\cong}{OP} }$  based on chosen initial beamforming vectors. The proposed algorithm to solve the original $OP$ is summarized in Algorithm 1.    
\begin{figure}[h]
    \hspace{0em}\hrulefill

 \hspace{0em} {\bf Algorithm 1:} SE-EE trade-off maximization using SCA

     \hspace{0em}\hrulefill
     
     \hspace{0em} Step 1: Check the feasibility of the problem

     \hspace{0em} Step 2: Initialization of $\Psi^{(0)}$

     \hspace{0em} Step 3: Repeat
\begin{enumerate}
   \item Solve the  optimization problem in (\ref{20}) 
   \item Update $ \Psi^{(n+1)} $ 
\end{enumerate}
\hspace{-0em} Step 4: Until required accuracy is achieved.

\hspace{-1em} \hrulefill
\end{figure} 
\subsubsection{Convergence Analysis} 

By making use of the   analysis presented in~\cite{sqa}, we provide the convergence analysis for the proposed algorithm.  Let us first indicate that the   optimization parameters at the n$^{th}$ iteration (i.e., $\Psi^{(n)})$ are updated based on the solution   obtained by solving the approximated optimization problem in (\ref{20}). To ensure the convergence of this algorithm, three key conditions have to  be satisfied. Firstly,  appropriate initial conditions are chosen to ensure the feasibility of the approximated problem $ {\overset{\cong}{OP} }$ at the first iteration of Algorithm~1. This provides a feasible solution to update the parameters in the  next iteration.   It is worth  mentioning  that the feasible solution to the approximated problem can always ensure the original constraint. In order to provide   an additional insight into  this feasibility issue, we include the following lemma:
\begin{lemma}    
Suppose that the   feasible solution set of the  optimization problem ${OP}$ is denoted by $\chi$; then, the feasible region of the approximated convex optimization problem   $\Psi^{n}$  falls within the same feasible region of the original non-convex problem, i.e.,  $ \Psi^{n}\subseteq \chi, \forall n$.
\end{lemma}

\textit{Proof}: To prove this lemma, we firstly  point out that the approximated  optimization problem $\overset{\cong}{OP}$ is solved iteratively. As such, at each iteration, the solution of $\overset{\approx}{OP}$ is provided for the given set of convex constraints in (39). Using the first-order Taylor series expansion,  these constraints in the original problem ${OP}$ are approximated with their lower bounds. This implies that   the solution  also lies within the same feasible region $ \Psi^{n}$  and    satisfies all the constraints in the original problem \cite{sqa}, i.e., $ \Psi^{n}\subseteq \chi$, which completes the proof of   Lemma 2. \hfill$\blacksquare$
\\ Secondly,   we  present   a new lemma     to support that the objective function in ${\overset{\cong}{OP} }$ is non-decreasing with each iteration.   
 \begin{lemma}    
 The objective function in $\overset{\cong}{OP}$ is non-decreasing in terms of $\Psi^{n}$,  i.e., $\Upsilon(\Psi^{(n+1)}) \geq \Upsilon(\Psi^{(n)})$, where $ \Upsilon(\Psi^{(n)})= \Gamma_1(\Psi^{(n)})+\Gamma_2(\Psi^{(n)})$.
\end{lemma} 
\textit{Proof}:  To prove this lemma, we point out that the solution of ${\overset{\cong}{OP} }$ at the $n^{th}$ iteration is   a feasible solution to  ${\overset{\cong}{OP} }$  in the next iteration.  This inherently means that
 the objective function  value at the $n^{th}$ iteration,   ${\Upsilon(\Psi^{(n)}})$, {  is less than or equal to that obtained in the subsequent iteration,}  $ {\Upsilon(\Psi^{(n+1)}})$, which means that ${\Upsilon(\Psi )}$ is non-decreasing function \cite{sqa}.  This completes  the proof of Lemma 3. \hfill$\blacksquare$   \\
    Therefore, the objective value at each iteration will either increase or remain the same.  Finally,  the power constraint   in (\ref{oper})  ensures that the objective function of $ {\overset{\cong}{OP} }$ is  upper bounded due to the fact that  $P_{ava} << \infty$. In particular, the satisfaction of these three conditions ensures that the developed SCA technique converges to a solution with a finite number of iterations.


\subsubsection{Performance Evaluation}
The solution of ${\overset{\sim}{OP} }$ is obtained by introducing   multiple slack variables and iteratively solving the problem with different approximations.  Hence,   the   performance evaluation of the proposed approach is important to assess its effectiveness.    As such, we compare the performance of the proposed Algorithm 1 with a benchmark scheme.    In particular, we use  the power minimization problem $OP_P$  as a benchmark scheme by reformulating it as a semi-definite programming (SDP) which  provides the optimal solution \cite{orreston} \cite{cuma}. In this SDP, we set the  rates that are achieved through  solving ${\overset{\sim}{OP} }$ as the minimum rate requirements for $OP_{P}$ and those achieved rates are denoted by  $R_i^{*}$, $\forall i \in \mathcal{K}$. Then, we set these rates  as minimum rate targets for $OP_P$.  Without loss of generality, with introducing new rank-one matrices $\mathbf{W}_i= \mathbf{w}_i^H \mathbf{w}_i$  and exploiting semi-definite relaxation,     the SDP form of $OP_P$ can be formulated with minimum rate constraints as  follows \cite{cuma8}: 
\begin{subequations}\label{eqn:sdp} 
\begin{align}
 {\overset{\sim}{OP_P}}:   
  \underset{\{\mathbf{W}_i\}_{i=1}^K }{\text{minimize }}
&   \sum_{i=1}^{K}\text{Tr}[\mathbf{W}_i] \\
  \text{subject to}~
&   \text{Tr}[\textbf{H}_k \mathbf{W}_i]-\eta_i^{*}\sum_{j=1}^{i-1}\text{Tr}[\textbf{H}_k \mathbf{W}_j] \geq \nonumber \\ &   \eta_i^{*}\sigma_k^2,  \forall i\in \mathcal{K}, k\leq i, \\
      &   \text{Tr}[\textbf{H}_i \mathbf{W}_1] \leq\text{Tr}[\textbf{H}_i \mathbf{W}_2]\leq \cdots   \nonumber \\ & \leq\text{Tr}[\textbf{H}_i \mathbf{W}_K],  \forall i \in \mathcal{K},\\
      &  \mathbf{W}_i=\mathbf{W}_i^H,\mathbf{W}_i\succeq0,  \forall i \in \mathcal{K},     
\end{align}
\end{subequations}
where $\eta_i^{*}=2^{R_i^{*}}-1$, while $\textbf{H}_i=\mathbf{h}_i \mathbf{h}_i^H $. The solutions (i.e., the beamforming vectors) achieved through solving  ${\overset{\sim}{OP_P}}$ are optimal  and will also be the solutions to the original problem $OP_P$ provided that they are rank-one matrices  \cite{fayzeh}, \cite{orreston}. 
Note that the beamforming vectors are determined through extracting the eigenvectors corresponding to the maximum eigenvalues of these rank-one matrices \cite{cumanan}. In particular, we demonstrate in the simulation results that the proposed SCA technique to solve ${\overset{\sim}{OP} }$   provides { \textit{approximately similar}} performance of  ${\overset{\sim}{OP_P}}$.

 \subsubsection{Complexity Analysis of the Proposed SCA Technique}

 Considering the fact that an iterative SCA algorithm is adopted to solve the original problem $OP$, it is crucial to define  the   computational complexity of the proposed algorithm. This can be achieved through determining the complexity associated with  solving the approximated convex optimization problem $\overset{\cong}{OP}$ at each iteration of this algorithm. In particular, at each iteration, a linear objective function (i.e., $\Gamma_1+\Gamma_2$) is optimized with a set of 
   SOC  and linear constraints, where the interior-point method  is employed to obtain the solution at each iteration \cite{point},  \cite{secondd}.  Therefore, the computational complexity at each iteration is primarily defined considering the complexity of obtaining the solution of such second-order cone programme (SOCP). In general, the  work required to solve an SOCP problem is at  most $\mathcal{O}(\mathcal{B}^2 \mathcal{V})$ \cite{secondd}, where $\mathcal{B}$ and $\mathcal{V}$ denote the number of optimization variables and the total dimensions of the SOCP optimization problem, respectively. Furthermore, an  iterative algorithm converges to a solution with an upper bound given as  $\mathcal{O}(\sqrt{\mathcal{C}} \log (\frac{1}{\epsilon}))$, where $\mathcal{C}$ and $\epsilon$ are the total number of constraints at each iteration and    the required accuracy, respectively. Now, with the developed SCA in hand, $\mathcal{B}$ and $\mathcal{V}$ are estimated as  
  $( 1.5{K^2} +4.5K +2NK+3+c)$ and $( 5.5{K^2} +5K +2NK+4+c)$, respectively, where $c$ is a constant related to the number of constraints that arise due to the relaxation of the exponential constraints in the interior-point method  \cite{cvx20}.   Furthermore, the total number of constraints $\mathcal{C}$ in  $\overset{\cong}{OP}$ is found to be    $( 2.5K^2+6.5K+6+c)$.\\

\section{Simulation Results}\label{sec4}
In this section, we provide simulation  results to support the effectiveness of the proposed SE-EE trade-off beamforming  design of a downlink MISO-NOMA  system  over the conventional designs. In particular, we   study the impact of the trade-off between the achieved EE and SE.
 In these simulations, we consider a base station   equipped with  three transmit antennas (i.e., $N=3$), which simultaneously transmits to five single-antenna  users  that are located at a distance of   1, 2, 3, 4, and 50 meters from the base station, respectively. The small-scale fading is chosen to be Rayleigh fading,   while  the path loss exponent  $\kappa$  and the noise variance of all users    $\sigma^2 $  are both set to be 1.  In addition,  the minimum SINR thresholds are set to be $ 10^{-2}$ for all the  users, i.e.,  $\eta_{th}=10^{-2}$.  The   amplifiers' gain $\epsilon_0$ is set to   $0.65$, whereas the power losses at the base station are assumed to be $40$ dBm (i.e., $P_l=40$ dBm). Furthermore,  we define  the available power at the base station by TX-SNR in dB, such that  $\text{TX-SNR (dB)}=10\log_{10}\frac{P_{ava}}{\sigma^{2}}. $ 
 In addition, the available bandwidth of transmission is assumed to be 1 MHz, i.e., $B_w=1$ MHz. Furthermore, the algorithm terminates when the difference between two consequent outputs is less than 0.001 (i.e., $\varepsilon\leq 0.01$).  Finally, we  define the achieved  sum rate of the cell as 
$$\text{sum rate}= B_w\text{SE}. $$
\begin{figure}[h]
\centering 
\includegraphics[width=.5\textwidth]{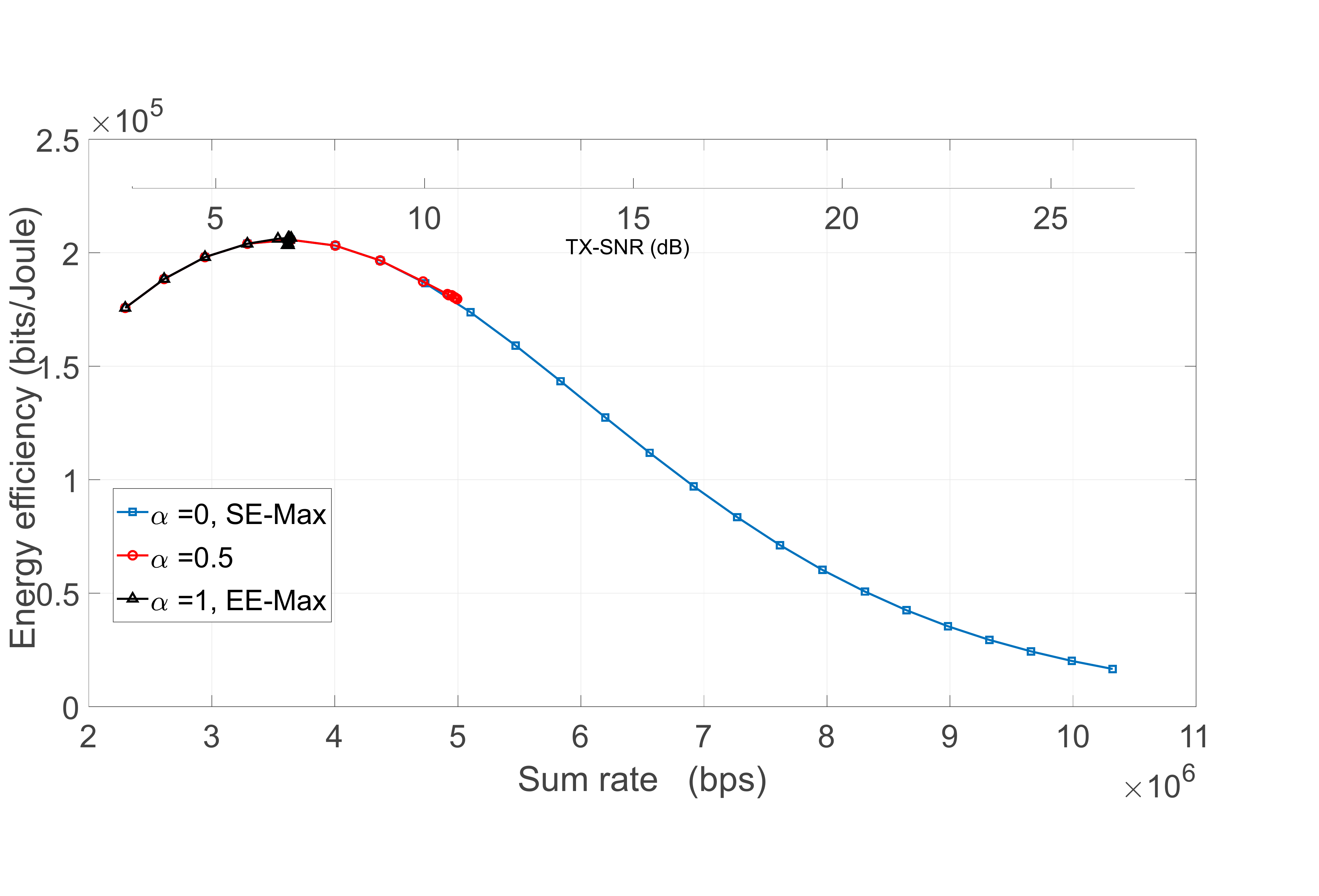}
\caption{Achieved EE and   sum rate   against TX-SNR with different   weight factors $\alpha$.}
\centering
\label{fig:rate}
\end{figure}

 Fig. \ref{fig:rate}  illustrates    the achieved EE and sum rate   versus different TX-SNR  and for different weight factors $\alpha$.   As seen in Fig. \ref{fig:rate},    the SE-EE trade-off design considered in   ${\overset{\sim}{OP} }$ turns out to be SE-Max with $\alpha=0$. Furthermore, ${\overset{\sim}{OP} }$ keeps maximizing the sum rate as   TX-SNR increases   at the cost of    EE degradation. This is due to the fact that the GEE (i.e., EE) has been assigned with zero weight (i.e., $\alpha =0$) in the MOO problem. However,  with  $\alpha=1$,   the problem is transformed into a GEE-Max design; as a result,  the maximum EE is achieved  with certain power threshold, referred  as green power in the literature. Beyond this green power, no further enhancement is achieved either in the EE  or  in the sum rate. Furthermore,  this design has the flexibility to strike a good balance between EE and the sum rate by setting the weight factor $\alpha$  between 0 and 1. { For example, when  $\alpha=0.5$}, an increment in the   sum rate is attained compared to that obtained with $\alpha=1$,  as seen in Fig. \ref{fig:rate}. However, this sum rate enhancement is attained at the cost of EE degradation.\\
 
\begin{figure}[!]
\centering
\includegraphics[width=.5\textwidth]{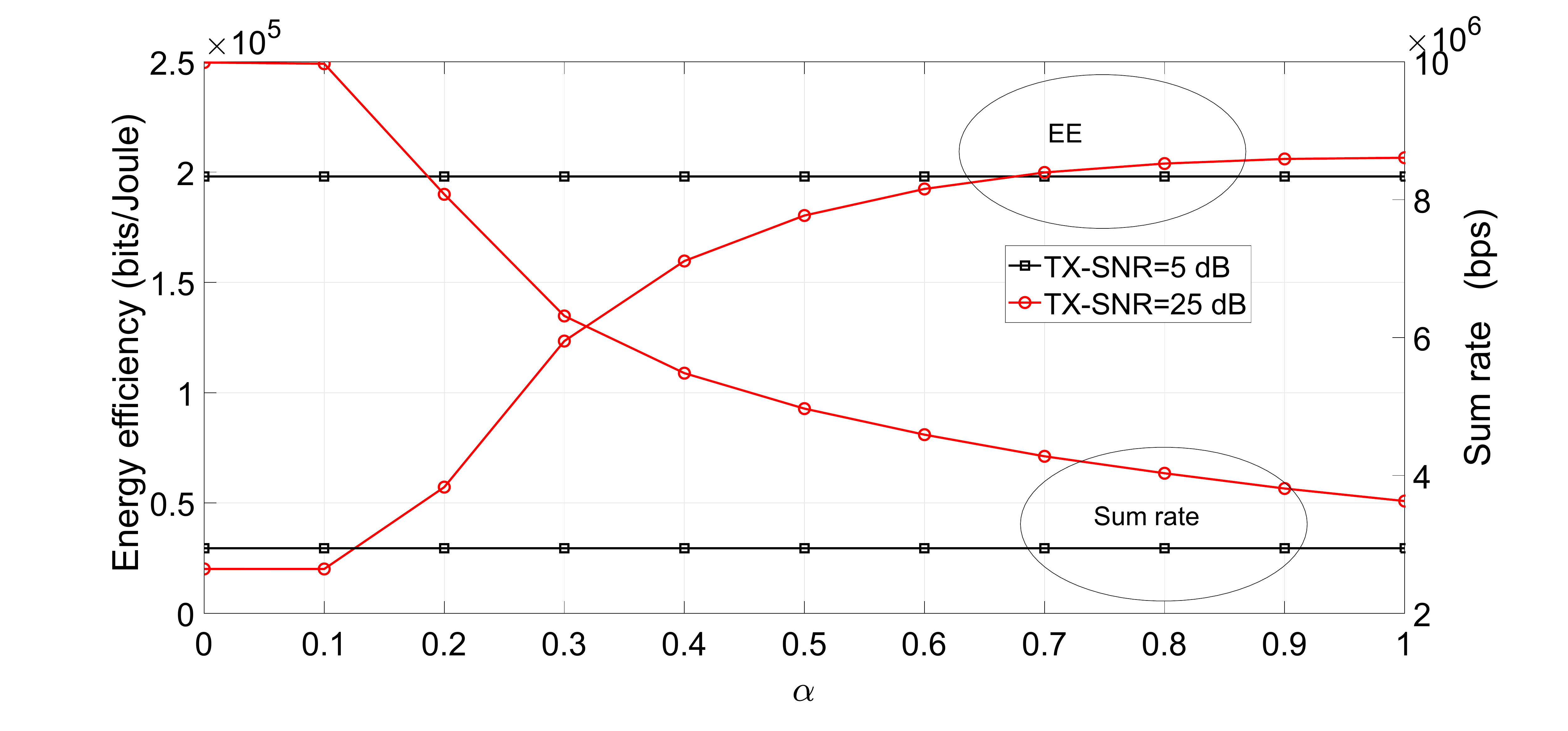}
\caption{Achieved EE and   sum rate    with different    weight factors $\alpha$. }
\centering
\label{fig:weight}
\end{figure}

 Fig. \ref{fig:weight} presents    the achieved EE and sum rate with different weight factors for 5 and 25 dB TX-SNR thresholds. In particular, Fig. \ref{fig:weight} {  shows} two different behaviors. First,  both performance metrics (i.e., sum rate and EE) remain constant with the available power lower than the green power for different  weight factors $\alpha$, which supports the validation of Lemma 1.   However, as TX-SNR exceeds the green power, for example with TX-SNR= 25 dB, the trade-off between EE and sum rate can be realized by varying the weight factor. In particular, at this TX-SNR threshold, the achieved rate and EE  show a performance in the range of   10-3.5 Mbps and 0.02-0.2 Mbits/Joule, respectively. This performance range is achieved with   different weight factors $\alpha$,  as  presented in  Fig. \ref{fig:weight}.    Note that the base station in the proposed SE-EE trade-off design offers a wide-range of SE-EE trade-off through a possibility of simply tuning the weigh factor $\alpha$. In fact, this  flexibility is  beneficial for different practical applications where the transmission techniques can  be adaptive according to the available power resources.

\begin{figure*} [htp]
\centering
\begin{subfigure}{.5\textwidth}
  \centering
  \includegraphics[width=1\linewidth]{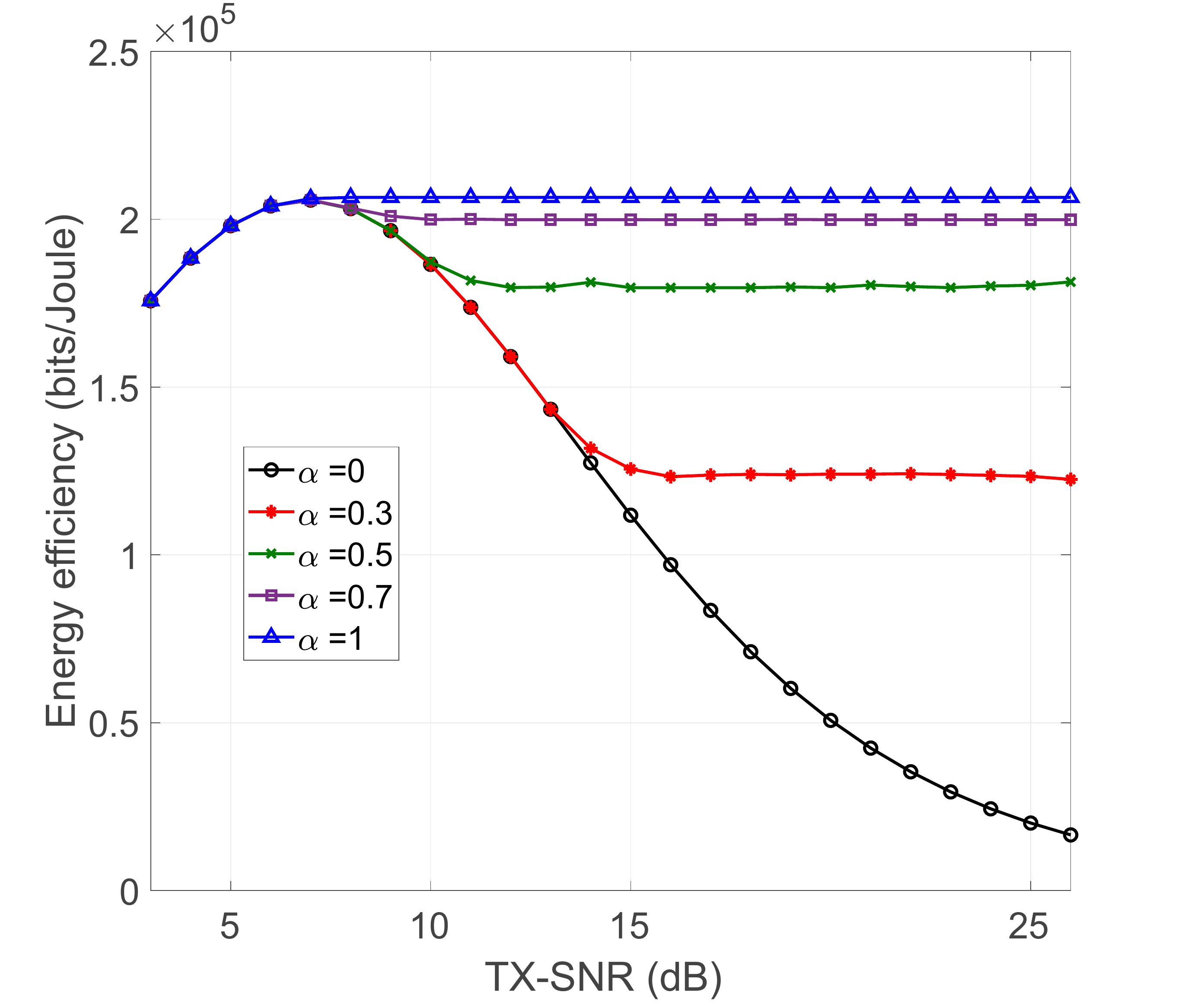}
   \caption{ }
  \label{fig:sub1}
\end{subfigure}%
\begin{subfigure}{.5\textwidth}
  \centering
  \includegraphics[width=1\linewidth]{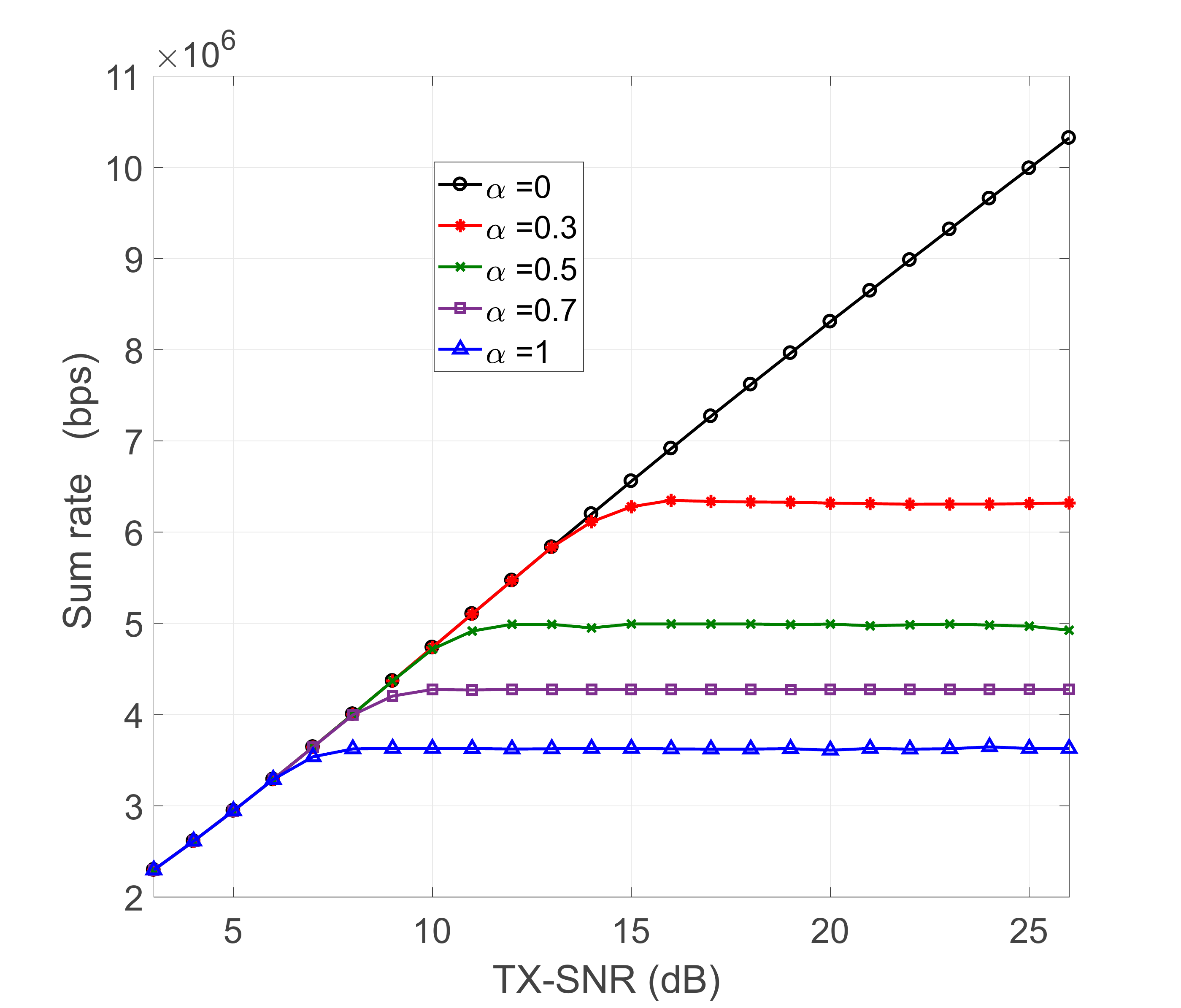}
   \caption{ }
  \label{fig:sub2}
\end{subfigure}
\caption{The EE and sum-rate performance of the proposed design  versus TX-SNR, with different weight factors. (a) The achieved EE, (b) the achieved sum rate.}
\label{fig:test}
\end{figure*}

Furthermore, Figs.~\ref{fig:sub1}  and~\ref{fig:sub2} show  the achieved EE and sum rate versus TX-SNR for different weight factors $\alpha$, respectively. It can be clearly understood   the impacts of the weight factors  on  the achieved EE and sum rate of the system. For example, at TX-SNR = 20 dB,   EE declines from  $ 2.2\times10^5$  bits/Joule   to $0.5\times10^5$ bits/Joule  by changing the weight factor  from  $\alpha=1$ to $\alpha=0$. However, the sum rate { (i.e., SE)} shows a different behavior. With TX-SNR = 20 dB, this decreases from  $8.2\times10^6$ bps to   $3.6\times10^6$ bps by changing  the weight factor $\alpha$   from 0 to 1. Therefore, the proposed EE-SE trade-off design offers the flexibility to the base station  to choose an appropriate   weight factor based on the favorable conditions and system requirements  to determine the desired performance metric.\\ 

{ 

\begin{table*}
\centering
\caption { Achieved EE and sum rate for the proposed design versus different TX-SNR, with different weight factors $\alpha$, $\eta_{th}= 1$.} 
\label{tab:powersinr1}
 \centering
    \begin{adjustbox}{width=\textwidth}
 \begin{tabular}{|c|c|c|c|c|c|c|c|c|c|}
    \hline
    \multirow{2}{*}{ } &
      \multicolumn{3}{c|}{TX-SNR= 5 dB}  &
      \multicolumn{3}{c|}{TX-SNR= 25 dB}  \\
      
      \cline{2-7}
    &sum rate (Mbps)& EE (Mbits/Joule) &P (W) & sum rate (Mbps)& EE (Mbits/Joule) & P (W) \\
    \hline
     $\alpha=0$ &  2.5301    &0.1702    & 3.1623  & 9.3011  & 0.0187  & 316.2278    \\
    \hline
     $\alpha=0.5$ & 2.5301    &0.1702   &  3.1623 & 6.4685 &      0.0635  & 59.7636  \\
    \hline
    $\alpha=1$  &2.5301    & 0.1702  &  3.1623  &  5.5084 & 0.0685 &  45.7811  \\
    \hline
  \end{tabular}
     \end{adjustbox}

\end{table*}

Furthermore, 	Fig. \ref{fig:NEW} demonstrates the impact of the minimum SINR threshold $\eta_{th}$ on the performance of the proposed design. In particular, with   $\eta_{th}=2$, the original optimization problem becomes infeasible as the minimum   required transmit power $P^*$   to achieve these minimum SINR requirements exceeds the available power $P_{ava}$. Hence, an alternative design, namely the SE-Max is considered. As a result, the achieved sum rate is maximized under the available power constraint which provides constant sum rate and EE over the different weight factors $\alpha$. However,  choosing lower value of $\eta_{th}$ ensures the feasibility of the design, which can be observed with   $\eta_{th}=0.2$, in Fig. \ref{fig:NEW}.    These results indicate that the base station has the flexibility  to choose either   EE  or SE performance metric based on the available energy resource by selecting an appropriate weight factor $\alpha$. Furthermore, the joint SE-EE design problem boils down to an SRM and GEE-max problem  with $\alpha = 0$ and $\alpha = 1$, respectively. 
\begin{figure}[H]\label{fig:NEW}
\centering
\includegraphics[width=.5\textwidth]{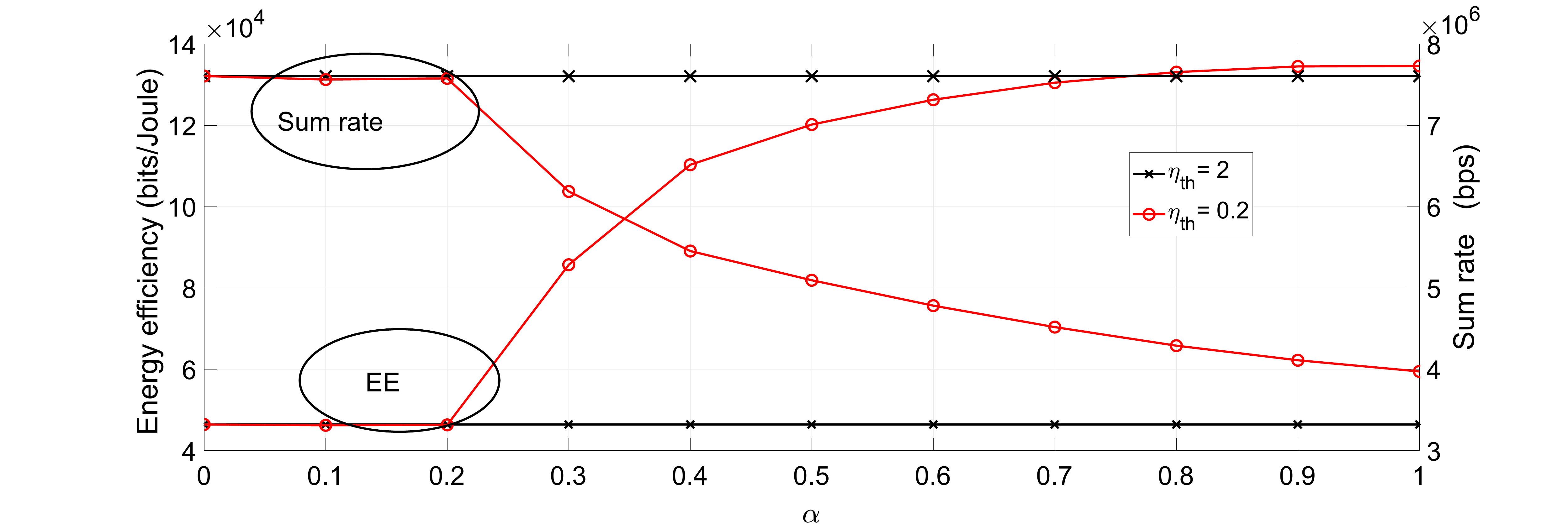}
\caption{Achieved EE and sum rate for the proposed design versus  weight factors $\alpha$, with different SINR thresholds $\eta_{th}$, TX-SNR= 20 dB.}
\centering
\label{fig:NEW}
\end{figure}

  In Table \ref{tab:powersinr2}, we show  the performance    of the proposed SCA algorithm. 
 The   rates achieved by solving  $\tilde{OP}$ (i.e., $R_i^*$) are set as target  SINR (i.e., $\eta_i^*$) for $\tilde{OP_P}$, where $\eta_i^*=2^{R_i^*}-1$. Then, the baseline optimization problem $\tilde{OP_P}$ is solved using the SDR approach.  In fact, through analysing the information in  Table \ref{tab:powersinr2},  we can   confirm that  the proposed SCA technique achieves  {approximately similar} solution to that of the benchmark, $OP_P$.

\begin{table*}
\centering
\label{tab:powersinr2}
\caption {Performance comparison between the proposed SCA algorithm to solve $\tilde{OP}$ and the benchmark solution obtained by solving $OP_P$  with $\eta_{th}= 0.2$ and TX-SNR= 20 dB.} \label{tab:powersinr2}
\begin{adjustbox}{width=\textwidth}
  \begin{tabular}{|l||l|l|l|l|l|l|l||l|l|    }
  \hline
    \multirow{2}{*}{ } &
      \multicolumn{7}{c||}{ $\tilde{OP}$}
      &\multicolumn{1}{c| }{ $\tilde{OP_P}$}\\
      \hline
      \cline{2-8}
    & $R_1^*$ (Mbps) & $R_2^*$ (Mbps)& $R_3^*$ (Mbps)& $R_4^*$ (Mbps)& $R_5^*$ (Mbps)& $R ^*$ (Mbps)& $P_t$ (W)& $P^*$ (W) \\
    \hline
    $\alpha=0.3$&0.4182    & 0.8068  &1.4025   & 1.6831   & 1.9579  &  6.2686 &43.2404   &43.1943 \\
    \hline
     $\alpha=0.5$& 0.4386   &0.8864   & 1.0159  &  1.4879  &1.5488   &  5.3776 &26.6188   &   26.2361
 \\
    \hline
    $\alpha=0.7$ & 0.4296   & 0.6662  &0.8142   &  1.3845  &  1.4474 &  4.7418 & 20.1294  &19.8267 \\
    \hline

  \end{tabular}
    \end{adjustbox}
\end{table*}

}

To further understand the impact of the TX-SNR in the feasibility of the optimization problem, and hence  on the EE-SE design, we provide the achieved sum rate and EE with different weight factors $\alpha$ in Table \ref{tab:powersinr1}. In particular, as observed in Table \ref{tab:powersinr1},  the minimum SINR threshold $\eta_{th}$ cannot be met when TX-SNR= 5 dB, hence, the EE-SE trade-off-based design becomes an SE-Max design. As such the sum rate is  maximized by solving $OP_{SE}$. Note that changing the weight factor $\alpha$ neither changes the sum rate nor the achieved EE.  However, with choosing TX-SNR= 25 dB, the minimum SINR threshold can be attained for this TX-SNR. Hence, the original optimization problem $OP$ is worthy to solve, and the achieved sum rate and EE for this case are presented in Table \ref{tab:powersinr1}.

\begin{figure}[H]\label{fig:nana1}
\centering
\includegraphics[width=.5\textwidth]{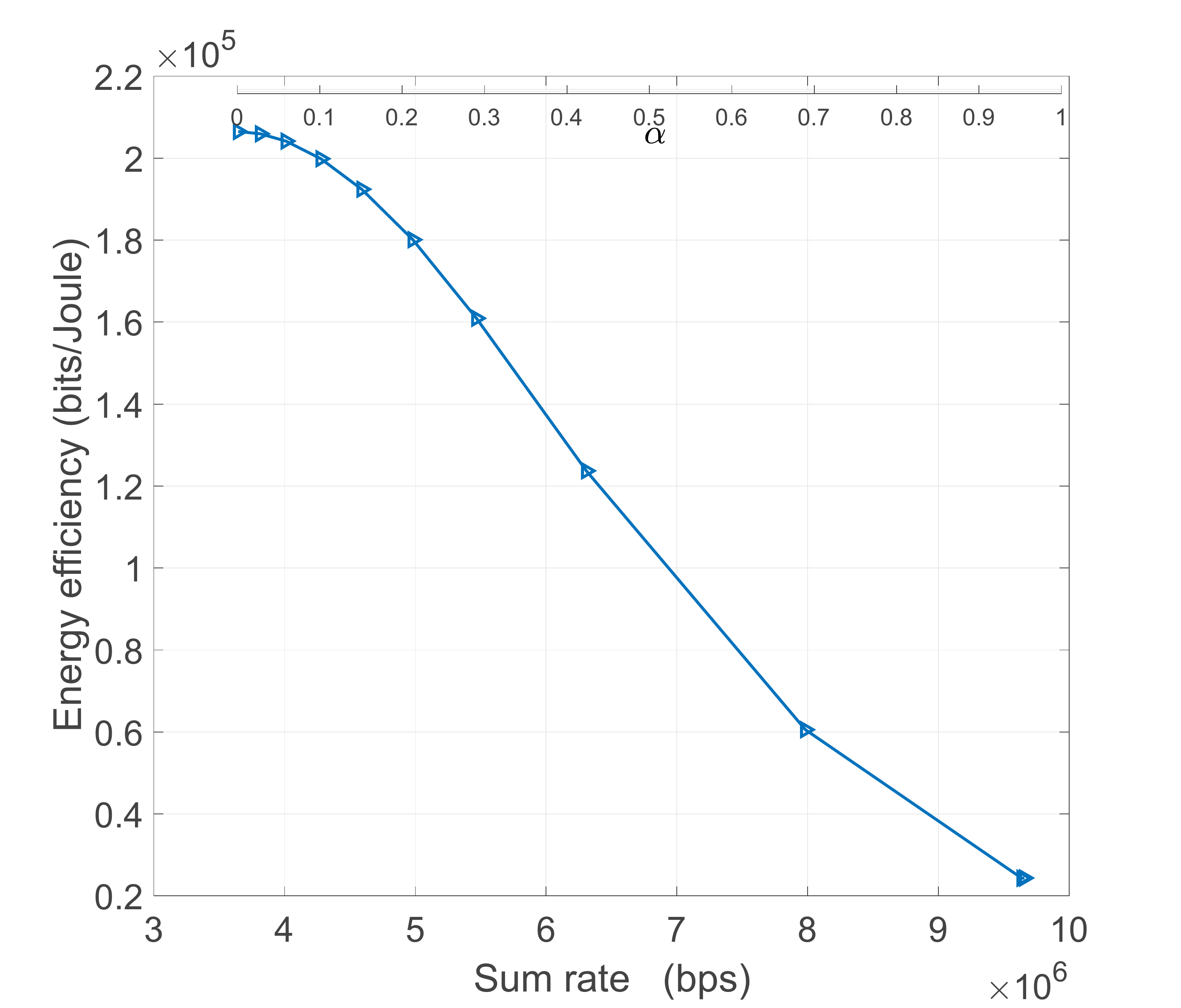}
\caption{Pareto front of SE-EE trade-off-based design for TX-SNR = 24 dB.}
\centering
\label{fig:paretof}
\end{figure}

Finally, Fig. \ref{fig:paretof} presents the set of Pareto-optimal solutions (i.e., Pareto-front) with TX-SNR = 24 dB. In particular, this curve provides all   best trade-off solutions (Pareto-optimal solutions) for the original SE-EE optimization problem. Furthermore, each point on this curve (sum rate and EE) corresponds to one of the best solutions that can be obtained with the corresponding weight factor. In other words,  any improvement in either one of the performance metrics with   a given weight factor can be only  achieved   by the degradation of the  other performance metric.

%

\section{Conclusion}\label{sec5}
In this paper, we proposed a  beamforming design that jointly  considers the maximization of the conflicting performance metrics  EE and SE. In particular, we formulate this challenging design problem through a  weighted sum approach based on the priori articulation. However, this original problem is not convex due to non-convex multi-objective function and constraints. To overcome these non-convexity issues, we exploit the  SCA technique to attain the solution. Furthermore, we show that the proposed approach  achieves a  Pareto-optimal solution. Simulation results have been provide to validate the effectiveness of the proposed algorithm and the performance is compared with a benchmark power minimization approach.

\section*{Appendix A} 
\section*{PROOF OF THEOREM 1}
First, we denote the  beamforming vectors that provide an optimal solution to  $\overset{\sim}{OP}$ as ${\{\mathbf{w}_i^*\}_{i=1}^{K}}$. Therefore, 
\begin{equation}
f_{EE-SE}({\{\mathbf{w}_i^*\}_{i=1}^{K}}) \geq f_{EE-SE}({\{\mathbf{w}_i\}_{i=1}^{K}}),
\end{equation}
which can be rewritten as 
\begin{equation}\label{eqa}
\sum_{l=1}^{2} \alpha_l f_l^{Norm}({\{\mathbf{w}_i^*\}_{i=1}^{K}})-\sum_{l=1}^{2} \alpha_l f_l^{Norm}({\{\mathbf{w}_i \}_{i=1}^{K}})   \geq 0.
\end{equation}
The inequality in (\ref{eqa}) can be equivalently reformulated as 
\begin{equation}\label{poo}
\sum_{l=1}^{2}  \frac{\alpha_l}{f_l^*} (f_l({\{\mathbf{w}_i^*\}_{i=1}^{K}})-f_l({\{\mathbf{w}_i \}_{i=1}^{K}}))\geq 0.
\end{equation}
In particular, we prove  Theorem 1 by using a contradiction argument,   as follows. First,  we assume that $ \{\mathbf{w}_i^*\}_{i=1}^{K}$ is not a  Pareto-optimal solution to the {  original optimization  problem ${OP}$.} This assumption implies that there exists   another feasible solution   $\{\mathbf{w}_i^{'}\}_{i=1}^{K} $ such that 
\begin{equation}\label{moko}
 \mathbf{f}{\{\mathbf{w}_i^{'}  \}_{i=1}^{K}} \succ \mathbf{f}{\{\mathbf{w}_i^*\}_{i=1}^{K}}.
\end{equation}
 The condition in (\ref{moko}) can be equivalently written as 
\begin{equation}\label{olp}
(f_l({\{\mathbf{w}_i^{'}\}_{i=1}^{K}})-f_l({\{\mathbf{w}_i^* \}_{i=1}^{K}}))> 0, \forall l \in {1,2}.
\end{equation}
Without loss of generality,   each element in (\ref{olp}) can be scaled by a positive constant (i.e., $ \frac{\alpha_l}{f_l^*}, \forall l \in \{1,2\} $). Furthermore, both of these inequalities can be added
\begin{equation}\label{hi}
\sum_{l=1}^{2}\frac{\alpha_l}{f_l^*}(f_l({\{\mathbf{w}_i^{'}\}_{i=1}^{K}})-f_l({\{\mathbf{w}_i^* \}_{i=1}^{K}}))> 0, \forall l \in {1,2}.
\end{equation}
However, the   inequality in (\ref{hi})  contradicts   the fact  that ${\{\mathbf{w}_i^*\}_{i=1}^{K}}$ is the optimal solution of  $\overset{\sim}{OP}$. Therefore, the optimal solution of  $\overset{\sim}{OP}$ 
  satisfies the Pareto-optimality condition,  and hence, it  gives the Pareto-solutions of the original SE-EE trade-off $OP$ problem. This completes the proof of Theorem 1. \hfill$\blacksquare$

\section*{Appendix B} 
\section*{PROOF OF LEMMA 1}
Lemma 1 presents that   $f_1^{Norm}$ and $f_2^{Norm}$ remain constant with   the different weight factors while the available power is less than green power. This can be equivalently written as 
\begin{subequations}
\begin{align}
\{f_1^{Norm}(\beta_1)\}_{P_{ava}=P_1}=\{f_1^{Norm}(\beta_2) \}_{P_{ava}=P_1}, \label{ok}\\
\{f_2^{Norm}(\beta_1)\}_{P_{ava}=P_1}=\{f_2^{Norm}(\beta_2) \}_{P_{ava}=P_1},\label{ok1}
\end{align}
\end{subequations}
where $P_1$ is less than the \text{green power}, and  $ \beta_1,\beta_2  \in [0,1]$.
In order to  prove this, we validate (\ref{ok}) and (\ref{ok1}) with the extreme conditions of  $\beta_1 =0$ \text{and} $\beta_2=1$.  We start with $\beta_1 =0$, in which case  ${\overset{\sim}{OP} }$ turns out to be an  SE-Max problem, and thus,  the maximum SE is achieved. Therefore, 
\begin{equation}\label{pl}
\{f_1^{Norm}(\beta_1=0)\}_{P_{ava}=P_1}=1.
\end{equation}
Furthermore, it has been already  { verified}  in \cite{haitham} that   both SE-Max and GEE-Max problem provide the same optimal beamforming vectors with an available power   less than the green power. This  means that $\{f_2 (\beta_1=0)\}_{P_{ava}=P_1}= f_2^{*} $, therefore, 
\begin{equation}\label{pl1}
\{f_2^{Norm}(\beta_1=0)\}_{P_{ava}=P1}=1.
\end{equation}
Similarly, we follow the same approach for the case with $\beta_2=1$, where   ${\overset{\sim}{OP} }$ becomes the GEE-Max problem. The maximization of EE with an available power   less than the green power  will simultaneously achieve the maximum sum rate and the maximum EE. Hence, 
\begin{subequations}
\begin{align}
\{f_1^{Norm}(\beta_2=1)\}_{P_{ava}=P_1}=1,\label{pl2}  \\
\{f_2^{Norm}(\beta_2=1)\}_{P_{ava}=P_1}=1.\label{pl3} 
\end{align}
\end{subequations}

It is can be easily noticed that   (\ref{pl}), (\ref{pl1}), (\ref{pl2}),  and (\ref{pl3}) validate the conditions provided in (\ref{ok}), (\ref{ok1}). This completes the proof of   Lemma 1.  \hfill$\blacksquare$

  \bibliography{references1}

\begin{thebibliography}{10}
\providecommand{\url}[1]{#1}
\csname url@samestyle\endcsname
\providecommand{\newblock}{\relax}
\providecommand{\bibinfo}[2]{#2}
\providecommand{\BIBentrySTDinterwordspacing}{\spaceskip=0pt\relax}
\providecommand{\BIBentryALTinterwordstretchfactor}{4}
\providecommand{\BIBentryALTinterwordspacing}{\spaceskip=\fontdimen2\font plus
\BIBentryALTinterwordstretchfactor\fontdimen3\font minus
  \fontdimen4\font\relax}
\providecommand{\BIBforeignlanguage}[2]{{%
\expandafter\ifx\csname l@#1\endcsname\relax
\typeout{** WARNING: IEEEtran.bst: No hyphenation pattern has been}%
\typeout{** loaded for the language `#1'. Using the pattern for}%
\typeout{** the default language instead.}%
\else
\language=\csname l@#1\endcsname
\fi
#2}}
\providecommand{\BIBdecl}{\relax}
\BIBdecl

\bibitem{book}
R.~Vannithamby and S.~Talwar, \emph{Towards 5G Applications: Requirements and
  Candidate Technologies}.\hskip 1em plus 0.5em minus 0.4em\relax John Wiley \&
  Sons, 2017.

\bibitem{green}
P.~Gandotra, R.~K. Jha, and S.~Jain, ``Green communication in next generation
  cellular networks: {A} survey,'' \emph{IEEE Access}, vol.~5, pp.
  11\,727--11\,758, Jun.~2017.

\bibitem{ee1}
A.~Zappone and E.~Jorswieck, ``Energy efficiency in wireless networks via
  fractional programming theory,'' \emph{Found. Trends Commun. Inf. Theory},
  vol.~11, no. 3-4, pp. 185--396, Jan.~2015.

\bibitem{massive1}
T.~L. Marzetta, ``Noncooperative cellular wireless with unlimited numbers of
  base station antennas,'' \emph{IEEE Trans. Wireless Commun.}, vol.~9, no.~11,
  pp. 3590--3600, Nov.~2010.

\bibitem{bashar2019energy}
M.~Bashar, K.~Cumanan, A.~G. Burr, H.~Q. Ngo, E.~G. Larsson, and P.~Xiao,
  ``Energy efficiency of the cell-free massive {MIMO} uplink with optimal
  uniform quantization,'' \emph{IEEE Trans. Green Commun. Netw.}, vol.~3,
  no.~4, pp. 971--987, Dec.~2019.

\bibitem{zhu2019joint}
L.~Zhu, J.~Zhang, Z.~Xiao, X.~Cao, D.~O. Wu, and X.-G. Xia, ``Joint {Tx-Rx}
  beamforming and power allocation for {5G} {Millimeter-Wave} non-orthogonal
  multiple access {(MmWave-NOMA) N}etworks,'' \emph{IEEE Trans. Commun.}, 2019.

\bibitem{xiao2018joint}
Z.~Xiao, L.~Zhu, J.~Choi, P.~Xia, and X.-G. Xia, ``Joint power allocation and
  beamforming for non-orthogonal multiple access {(NOMA)} in {5G} millimeter
  wave communications,'' \emph{IEEE Trans. Wireless Commun.}, vol.~17, no.~5,
  pp. 2961--2974, May~2018.

\bibitem{noma2}
S.~Tomida and K.~Higuchi, ``Non-orthogonal access with {SIC} in cellular
  downlink for user fairness enhancement,'' in \emph{Proc. IEEE Inter. Symp. on
  Intell. Signal Process. and Comm. Sys. (ISPACS), 2011}, 2011, pp. 1--6.

\bibitem{fnoma}
Y.~Saito, Y.~Kishiyama, A.~Benjebbour, T.~Nakamura, A.~Li, and K.~Higuchi,
  ``Non-orthogonal multiple access {(NOMA)} for cellular future radio access,''
  in \emph{Proc. IEEE VTC Spring 2013}, 2013, pp. 1--5.

\bibitem{noma5}
Z.~Ding, Y.~Liu, J.~Choi, Q.~Sun, M.~Elkashlan, I.~Chih-Lin, and H.~V. Poor,
  ``Application of non-orthogonal multiple access in {LTE} and {5G} networks,''
  \emph{IEEE Commun. Mag.}, vol.~55, no.~2, pp. 185--191, Feb.~2017.

\bibitem{noma3}
S.~Vanka, S.~Srinivasa, Z.~Gong, P.~Vizi, K.~Stamatiou, and M.~Haenggi,
  ``Superposition coding strategies: Design and experimental evaluation,''
  \emph{IEEE Trans. Wireless Commun.}, vol.~11, no.~7, pp. 2628--2639,
  Jul.~2012.

\bibitem{cuma9}
P.~Xu, K.~Cumanan, and Z.~Yang, ``Optimal power allocation scheme for {NOMA}
  with adaptive rates and alpha-fairness,'' in \emph{Proc. IEEE GLOBECOM},
  2017, pp. 1--6.

\bibitem{crnoma}
L.~Lv, J.~Chen, and Q.~Ni, ``Cooperative non-orthogonal multiple access in
  cognitive radio,'' \emph{IEEE Commun. Lett.}, vol.~20, no.~10, pp.
  2059--2062, Oct.~2016.

\bibitem{ref1}
B.~Wang, L.~Dai, Z.~Wang, N.~Ge, and S.~Zhou, ``Spectrum and energy-efficient
  beamspace {MIMO-NOMA} for millimeter-wave communications using lens antenna
  array,'' \emph{IEEE J. Sel. Areas Commun.}, vol.~35, no.~10, pp. 2370--2382,
  Oct.~2017.

\bibitem{ref3}
Z.~Wei, L.~Zhao, J.~Guo, D.~W.~K. Ng, and J.~Yuan, ``A multi-beam {NOMA}
  framework for hybrid mmwave systems,'' in \emph{Proc. IEEE International
  Conference on Communications (ICC)}, 2018, pp. 1--7.

\bibitem{mimonom2}
Q.~Sun, S.~Han, I.~Chin-Lin, and Z.~Pan, ``On the ergodic capacity of {MIMO
  NOMA} systems,'' \emph{IEEE Wireless Commun. Lett.}, vol.~4, no.~4, pp.
  405--408, Aug.~2015.

\bibitem{misonoma}
Z.~Chen, Z.~Ding, P.~Xu, and X.~Dai, ``Optimal precoding for a {Q}o{S}
  optimization problem in two-user {MISO-NOMA} downlink,'' \emph{IEEE Commun.
  Lett.}, vol.~20, no.~6, pp. 1263--1266, Jun.~2016.

\bibitem{zhang2019}
M.~Zhang, K.~Cumanan, J.~Thiyagalingam, W.~Wang, A.~G. Burr, Z.~Ding, and O.~A.
  Dobre, ``Energy efficiency optimization for secure transmission in {MISO}
  cognitive radio network with energy harvesting,'' \emph{IEEE Access}, vol.~7,
  pp. 126\,234--126\,252, 2019.

\bibitem{xin}
X.~Wei, H.~Al-Obiedollah, K.~Cumanan, M.~Zhang, J.~Tang, W.~Wang, and O.~A.
  Dobre, ``Resource allocation technique for hybrid {TDMA-NOMA} system with
  opportunistic time assignment,'' \emph{arXiv preprint arXiv:2003.02196},
  2020.

\bibitem{mimonoma}
Z.~Ding, F.~Adachi, and H.~V. Poor, ``The application of {MIMO} to
  non-orthogonal multiple access,'' \emph{IEEE Trans. Wireless Commun.},
  vol.~15, no.~1, pp. 537--552, Jan.~2016.

\bibitem{sun2018robust}
H.~Sun, F.~Zhou, R.~Q. Hu, and L.~Hanzo, ``Robust beamforming design in a
  {NOMA} cognitive radio network relying on {SWIPT},'' \emph{IEEE Sel. Areas in
  Commun.}, vol.~37, no.~1, pp. 142--155, Jan.~2019.

\bibitem{zhou2018artificial}
F.~Zhou, Z.~Chu, H.~Sun, R.~Q. Hu, and L.~Hanzo, ``Artificial noise aided
  secure cognitive beamforming for cooperative {MISO-NOMA} using {SWIPT},''
  \emph{IEEE Sel. Areas in Commun.}, vol.~36, no.~4, pp. 918--931, Apr.~2018.

\bibitem{zhou2017robust}
F.~Zhou, Z.~Li, J.~Cheng, Q.~Li, and J.~Si, ``Robust {AN}-aided beamforming and
  power splitting design for secure {MISO} cognitive radio with {SWIPT},''
  \emph{IEEE Trans Wireless Commun.}, vol.~16, no.~4, pp. 2450--2464,
  Apr.~2017.

\bibitem{liu2018non}
Y.~Liu, Z.~Qin, M.~Elkashlan, Z.~Ding, A.~Nallanathan, and L.~Hanzo,
  ``Non-orthogonal multiple access for {5G} and beyond,'' \emph{arXiv preprint
  arXiv:1808.00277}, 2018.

\bibitem{alavi2017outage}
F.~Alavi, K.~Cumanan, Z.~Ding, and A.~G. Burr, ``Outage constraint based robust
  beamforming design for non-orthogonal multiple access in {5G} cellular
  networks,'' in \emph{Proc. IEEE 28th Annual International Symposium on
  Personal, Indoor, and Mobile Radio Communications (PIMRC)}, 2017, pp. 1--5.

\bibitem{hanif}
M.~F. Hanif, Z.~Ding, T.~Ratnarajah, and G.~K. Karagiannidis, ``A
  minorization-maximization method for optimizing sum rate in the downlink of
  non-orthogonal multiple access systems,'' \emph{IEEE Trans. Signal Process.},
  vol.~64, no.~1, pp. 76--88, Jan.~2016.

\bibitem{haitham}
H.~M. {Al-Obiedollah}, K.~Cumanan, J.~Thiyagalingam, A.~G. Burr, Z.~Ding, and
  O.~A. Dobre, ``Energy efficient beamforming design for {MISO} non-orthogonal
  multiple access systems,'' \emph{IEEE Trans. Commun.}, vol.~67, no.~6, pp.
  4117--4131, Jun.~2019.

\bibitem{wcnc1}
H.~Al-Obiedollah, K.~Cumanan, J.~Thiyagalingam, A.~G. Burr, Z.~Ding, and O.~A.
  Dobre, ``Energy efficiency fairness beamforming design for {MISO} {NOMA}
  systems,'' in \emph{Proc. IEEE WCNC}, 2019.

\bibitem{energyhybrid}
J.~Lorincz and I.~Bule, ``Renewable energy sources for power supply of base
  station sites,'' \emph{International Journal of Business Data Communications
  and Networking (IJBDCN)}, vol.~9, no.~3, pp. 53--74, 2013.

\bibitem{wang2012survey}
L.-C. Wang and S.~Rangapillai, ``A survey on green {5G} cellular networks,'' in
  \emph{Proc. International Conference on Signal Processing and Communications
  (SPCOM)}, 2012, pp. 1--5.

\bibitem{xiong2011energy}
C.~Xiong, G.~Y. Li, S.~Zhang, Y.~Chen, and S.~Xu, ``Energy-and
  spectral-efficiency tradeoff in downlink {OFDMA} networks,'' \emph{IEEE
  Trans. Wireless Commun.}, vol.~10, no.~11, pp. 3874--3886, Dec.~2011.

\bibitem{alavi2019}
F.~Alavi, K.~Cumanan, M.~Fozooni, Z.~Ding, S.~Lambotharan, and O.~A. Dobre,
  ``Robust energy-efficient design for miso non-orthogonal multiple access
  systems,'' \emph{IEEE Transactions on Communications}, vol.~67, no.~11, pp.
  7937--7949, 2019.

\bibitem{eese}
L.~Deng, Y.~Rui, P.~Cheng, J.~Zhang, Q.~Zhang, and M.~Li, ``A unified energy
  efficiency and spectral efficiency tradeoff metric in wireless networks,''
  \emph{IEEE Commun. Lett.}, vol.~17, no.~1, pp. 55--58, Jan.~2013.

\bibitem{MO1}
A.~Konak, D.~W. Coit, and A.~E. Smith, ``Multi-objective optimization using
  genetic algorithms: A tutorial,'' \emph{Reliability Engineering \& System
  Safety}, vol.~91, no.~9, pp. 992--1007, Sep.~2006.

\bibitem{MO2}
A.~Zhou, B.-Y. Qu, H.~Li, S.-Z. Zhao, P.~N. Suganthan, and Q.~Zhang,
  ``Multiobjective evolutionary algorithms: A survey of the state of the art,''
  \emph{Swarm and Evolutionary Computation}, vol.~1, no.~1, pp. 32--49,
  Mar.~2011.

\bibitem{MO3}
R.~T. Marler and J.~S. Arora, ``Survey of multi-objective optimization methods
  for engineering,'' \emph{Structural and Multidisciplinary Optimization},
  vol.~26, no.~6, pp. 369--395, Apr.~2004.

\bibitem{wcnc2}
H.~Al-Obiedollah, K.~Cumanan, J.~Thiyagalingam, A.~G. Burr, Z.~Ding, and O.~A.
  Dobre, ``Sum rate fairness trade-off-based resource allocation technique for
  {MISO NOMA} systems,'' in \emph{Proc. IEEE WCNC}, 2019.

\bibitem{eese1}
O.~Amin, E.~Bedeer, M.~H. Ahmed, and O.~A. Dobre, ``Energy efficiency--spectral
  efficiency tradeoff: A multiobjective optimization approach,'' \emph{IEEE
  Trans. Veh. Technol.}, vol.~65, no.~4, pp. 1975--1981, Apr.~2016.

\bibitem{octavia}
E.~Bedeer, O.~A. Dobre, M.~H. Ahmed, and K.~E. Baddour, ``A multiobjective
  optimization approach for optimal link adaptation of {OFDM}-based cognitive
  radio systems with imperfect spectrum sensing,'' \emph{IEEE Trans. Wireless
  Commun.}, vol.~13, no.~4, pp. 2339--2351, Apr.~2014.

\bibitem{zhang2015energy}
W.~Zhang, C.-X. Wang, D.~Chen, and H.~Xiong, ``Energy--spectral efficiency
  tradeoff in cognitive radio networks,'' \emph{IEEE Trans. Veh. Technol.},
  vol.~65, no.~4, pp. 2208--2218, May 2015.

\bibitem{hong2013energy}
X.~Hong, Y.~Jie, C.-X. Wang, J.~Shi, and X.~Ge, ``Energy-spectral efficiency
  trade-off in virtual {MIMO} cellular systems,'' \emph{IEEE Sel. Areas in
  Commun.}, vol.~31, no.~10, pp. 2128--2140, Sep.~2013.

\bibitem{itu12}
G.~Liu and D.~Jiang, ``{5G}: Vision and requirements for mobile communication
  system towards year 2020,'' \emph{Chinese Journal of Engineering}, vol. 2016,
  Mar.~2016,~Art. no. 5974586.

\bibitem{fayzeh}
F.~Alavi, K.~Cumanan, Z.~Ding, and A.~G. Burr, ``Beamforming techniques for
  non-orthogonal multiple access in {5G} cellular networks,'' \emph{IEEE Trans.
  Veh. Technol.}, vol.~67, no.~10, pp. 9474--9487, Oct.~2018.

\bibitem{cuma2}
P.~Xu and K.~Cumanan, ``Optimal power allocation scheme for non-orthogonal
  multiple access with $\alpha $-fairness,'' \emph{IEEE J. Sel. Areas in
  Commun.}, vol.~35, no.~10, pp. 2357--2369, Oct.~2017.

\bibitem{noma4}
S.~R. Islam, N.~Avazov, O.~A. Dobre, and K.-S. Kwak, ``Power-domain
  non-orthogonal multiple access {(NOMA)} in {5G} systems: {P}otentials and
  challenges,'' \emph{IEEE Commun. Surveys Tuts.}, vol.~19, no.~2, pp.
  721--742, Oct.~2017.

\bibitem{mojtaba}
M.~Vaezi, R.~Schober, Z.~Ding, and H.~V. Poor, ``Non-orthogonal multiple
  access: Common myths and critical questions,'' \emph{IEEE Wireless
  Communications}, vol.~26, no.~5, pp. 174--180, 2019.

\bibitem{zhang2017secure}
M.~Zhang, K.~Cumanan, and A.~Burr, ``Secure energy efficiency optimization for
  {MISO} cognitive radio network with energy harvesting,'' in \emph{Proc. IEEE
  9th International Conference on Wireless Communications and Signal Processing
  (WCSP)}, 2017, pp. 1--6.

\bibitem{oct4}
M.~Zeng, A.~Yadav, O.~A. Dobre, and H.~V. Poor, ``Energy-efficient power
  allocation for {MIMO-NOMA} with multiple users in a cluster,'' \emph{IEEE
  Access}, vol.~6, pp. 5170--5181, Feb.~2018.

\bibitem{islam2018}
S.~Islam, M.~Zeng, O.~A. Dobre, and K.-S. Kwak, ``Resource allocation for
  downlink {NOMA} systems: Key techniques and open issues,'' \emph{IEEE
  Wireless Commun.}, vol.~25, no.~2, pp. 40--47, Apr.~2018.

\bibitem{sqa}
A.~Beck, A.~Ben-Tal, and L.~Tetruashvili, ``A sequential parametric convex
  approximation method with applications to nonconvex truss topology design
  problems,'' \emph{J. Global Optimiz.}, vol.~47, no.~1, pp. 29--51, May~2010.

\bibitem{harv}
H.~Al-Obiedollah, K.~Cumanan, A.~G. Burr, J.~Tang, Y.~Rahulamathavan, Z.~Ding,
  and O.~A. Dobre, ``On energy harvesting of hybrid {TDMA-NOMA} systems,'' in
  \emph{Proc. IEEE Global Communications Conference (GLOBECOM)}, 2019, pp.
  1--6.

\bibitem{convex}
S.~Boyd and L.~Vandenberghe, \emph{Convex {O}ptimization}.\hskip 1em plus 0.5em
  minus 0.4em\relax Cambridge Univ. Press, 2004.

\bibitem{orreston}
M.~Bengtsson and B.~Ottersten, ``Optimal downlink beamforming using
  semidefinite optimization,'' in \emph{Proc. Annual Allerton Conf. on Commun.,
  Control and Computing}, 1999, pp. 987--996.

\bibitem{cuma}
K.~Cumanan, L.~Musavian, S.~Lambotharan, and A.~B. Gershman, ``Sinr balancing
  technique for downlink beamforming in cognitive radio networks,'' \emph{IEEE
  Signal Processing Letters}, vol.~17, no.~2, pp. 133--136, 2010.

\bibitem{cuma8}
F.~Alavi, K.~Cumanan, Z.~Ding, and A.~G. Burr, ``Robust beamforming techniques
  for non-orthogonal multiple access systems with bounded channel
  uncertainties,'' \emph{IEEE Commun. Lett.}, vol.~21, no.~9, pp. 2033--2036,
  Sept.~2017.

\bibitem{cumanan}
K.~Cumanan, R.~Krishna, L.~Musavian, and S.~Lambotharan, ``Joint beamforming
  and user maximization techniques for cognitive radio networks based on branch
  and bound method,'' \emph{IEEE Trans. Wireless Commun.}, vol.~9, no.~10, pp.
  3082--3092, Oct.~2010.

\bibitem{point}
Y.~Nesterov and A.~Nemirovskii, \emph{Interior-point Polynomial Algorithms in
  Convex Programming}.\hskip 1em plus 0.5em minus 0.4em\relax Philadelphia, PA:
  SIAM, 1994.

\bibitem{secondd}
M.~S. Lobo, L.~Vandenberghe, S.~Boyd, and H.~Lebret, ``Applications of
  second-order cone programming,'' \emph{Linear Algebra and its Appl.}, vol.
  284, pp. 193--228, Nov.~1998.

\bibitem{cvx20}
M.~Grant, S.~Boyd, and Y.~Ye, ``{CVX}: Matlab software for disciplined convex
  programming,'' 2008.

\end{thebibliography}
  \bibliographystyle{ieeetran}

\end{document}